\documentclass{emulateapj}

\usepackage[T1]{fontenc}
\usepackage{ae,aecompl}
\usepackage{natbib}
\usepackage{wasysym}

\usepackage{xcolor}
\begin{document}
\title{Dynamically Selected Mass-Radius Relationship for Low Mass  Exoplanets}
\author{Brad M. S. Hansen}
\email[email]{ hansen@astro.ucla.edu}
\affil{ Mani L. Bhaumik Institute for Theoretical Physics \\  Department of Physics \& Astronomy, University of California Los Angeles, \\ Los Angeles, CA 90095}
\author{Austin Mao}
\affil{ Department of Physics \& Astronomy, University of California Los Angeles, \\ Los Angeles, CA 90095}

\begin{abstract}

We study the mass--radius relationship of planets in multiplanet systems as a function of their dynamical architecture. We isolate those planets whose proximity to resonance
indicates that they have not undergone significant dynamical instability since formation, and therefore have not experienced a late giant
impact with another planet. We compare the properties of these planets to those whose orbital architectures suggest a high probability
of having experienced a giant impact.

We find that planets that are likely to have experienced an impact are, on average, more massive than those that did not -- consistent with
prior claims. However, we find that the Hydrogen envelope mass fractions of these planets are no smaller than those of planets that did not
experience collisions. Taken together, these findings suggest that dynamical evolution and planetary collisions are an integral part of the evolution
of compact planetary systems, but that they must occur early enough that collisional remnants are still able to recapture gaseous envelopes
from the dissipating protoplanetary disk.

 \end{abstract}

\maketitle


\section{Introduction}

The discovery of extrasolar planets around other stars has demonstrated that our own Solar system planets span only a fraction of
the available parameter space \cite[e.g.][]{WF15,WMP23}. In particular, a significant fraction of Sun-like stars are orbited by multiple planets whose orbital periods are
significantly shorter than any of those in the Solar system. The origin of these planets is a matter of active debate, as the mass inventory
of solid material required is significantly larger than a naive protoplanetary disk model would imply \citep{CL13,DBL23}. While it seems clear that solid
material must condense on larger scales and then migrate inwards, it is debated as to whether this material migrates as smaller bodies,
due to aerodynamic drag on the gas disk \citep{W77},
that later assemble in place \citep{HM12,HM13,CT14}, or assembles into larger bodies and then migrates inwards due to torques that exchange angular
momentum with the gas disk \citep{IL08,IL10,KN12,PDD23}.

The physical structure of these planets is also debated. Measurements of both the mass and radius can yield crude constraints on the bulk
composition of the planets. Some planets are consistent with a rocky composition, but many have radii too large to be explained by a purely
silicate mass-radius relation \citep{GPD07,CBI09,ERC12,LJR13}. Moderately enhanced radii can be explained with a thick water envelope that overlies a rocky core, while the
largest radii require the presence of a Hydrogen-rich envelope \citep{LF14}, of at least several percent by mass (much larger than for the Solar system
terrestrial planets). The origin of planets with  low density envelopes is also influenced by the birth location conditions of these
planets,  and by their subsequent migration  history.
So these two foundational questions for exoplanet origins are strongly coupled.

Many models for the formation of compact, exoplanet systems invoke migration inwards of planets and protoplanets from larger scales due
to the launching of waves in the protoplanetary disk \citep{GT80}.  In this model, individual planets migrate quite rapidly, but get trapped into mean motion
resonances, which slows the migration and allows for the planets to survive -- especially if the stellar magnetic field maintains an interior
cavity to the disk (so that the planetary migration does not carry the planets into the star). The natural outcome of this process is to produce
a set of planets locked in a chain of mean motion resonances \citep{IL08,IL10}. However, the observed population does not contain a significant fraction of
resonant nearest neighbour pairs \citep{LRF11,FLR14}.

Thus, the principal difference among contemporary iterations of this model centers on how the planets break out of their initially resonant
configurations. In principle, a chain of planets in a very compact resonant configuration would be unstable to chaotic eccentricity
excitation due to the overlapping of neigbouring resonances \citep{Wis80,Q11}, which would lead to dynamical instability. Such chains remain
stable as long as they are embedded in the protoplanetary disk because of the eccentricity damping by the gas. However, the 
eventual dispersal of the disk removes this stabilising influence.
 The resulting eccentricity excitation leads to orbit crossing and planetary collisions,
reducing the number of planets, and producing more widely spaced systems that are dynamically stable \citep{IOR17,IBR21}. This also has potential
implications for the observed planetary mass--radius relations, because the lower density envelopes are susceptible to being lost during
the violent episodes of a giant impact \citep{LHLA15,IS16,BS19}.

An alternative route to the breaking of resonant chains is to model the outward movement of the disk inner edge as the accretion rate
drops and the magnetospheric cavity grows \citep{LOL17}. For a sharp inner edge, this model breaks some of the resonant locks, but
not enough to match the observations \citep{LO17}. However, a more detailed treatment of the magnetic field diffusion into the disk
results in a more gradual rise to a surface density maximum located exterior to the true magnetospheric cavity \citep{YHH23}. 
Incorporating this model into the magnetospheric rebound model promotes divergent migration of resonant pairs as the
 magnetospheric
cavity expands, and results in the majority of the resonant pairs being broken \citep{HYH24,HYNH25} during the final dissipation
of the protoplanetary disk. This is a fundamentally different
pathway to match the observations, because it requires that fewer of the planets experience large scale impacts \citep{HYNH25}, and most
of those that do occur happen during the earlier phase when some of  the gas disk is still present.

A third, intermediate, model invokes scattering off a population of leftover planetesimals \citep{CF15,WML24} to drive neighbouring planets apart,
thereby breaking the resonant lock and also promoting large scale dynamical instability. For the observed population of planets, the ultimate fate of the scattered planetesimals is accretion by
the planets involved, so this will also have an effect on the planetary mass inventory, but the consequences may not be as violent as in the
case of a single impact by a planet-mass body. Furthermore, the timescale on which dynamical instability is induced will depend on the unknown masses of the scattering bodies, with smaller bodies promoting earlier instability \citep{HW26}.

These different paths from formation to observation make somewhat different predictions as to the collisional histories of the
planets that remain in the chains. In the magnetospheric rebound models \citep[e.g.][]{HYNH25}, most planets either retain the envelopes they
accreted and carried through the nebular stage, or experience a collision early enough that they may potentially replace
some of the lost atmosphere from the ambient nebula.  For
the  models which rely on late-time dynamical instabilities \citep[e.g.][]{IBR21}, most surviving planets should be the outcomes of
planetary scale collisions and mergers taking place without any remaining ambient gas, and their contents should reflect the heating and mass ejection expected in such events.
 The planetesimal scattering models \citep[e.g.][]{CF15} predict that planets should have
accreted a few percent of their mass in additional rocky bodies, but may not have lost significant volatile content in giant impacts. They may
also experience dynamical instabilities, but the precise timing depends on the properties of the scattering bodies.

Indeed, initial observations  comparing the properties of Neptune-size planets, as a function of proximity to resonance, indicate that
those farther from resonance are higher in mass, and of higher density, than those closer to resonance \citep{LDB24,LCCD25}. This
intuitively conforms to the idea that planets farther from resonance have lost mass due to giant impacts.
However, this naive impression does not account for the compressibility of planets -- models show that more massive planets, at
fixed Hydrogen mass fraction, also have higher mean density \citep{LF14}. Furthermore, in the magnetospheric rebound models
of \citep{HYH24,HYNH25}, higher mass planets remaining coupled to the gas disk for longer, and therefore 
diverge farther from resonance, even in the absence of collisions.

Therefore, it is the goal of this paper to study these correlations. We wish to examine the properties of the planetary mass--radius relation as a function of the dynamical architecture 
of the planetary systems. We aim to identify any structural differences in planets that remain in resonant configurations, versus those
that do not. In \S~\ref{Boundary} we perform numerical simulations of dynamically unstable resonant chains in order to define a
quantitative boundary between those planets that remain in primordial resonances, versus those that do not, and how that correlates
with the collisional history of the final planets. In \S~\ref{DynMR} we then use these results to classify planets according to their
observed dynamical properties, 
 and to examine the resulting empirical mass--radius relationships. We   discuss the implications in \S~\ref{Infer}.

\section{Defining the Dynamical Boundary}
\label{Boundary}

A necessary condition for  a mean motion resonance between two planets is that the ratio of their orbital periods should be close to a rational number.
The technical definition of resonance requires that a resonant angle defined by that period ratio librate rather than circulate, but we do not usually
acquire sufficient orbital information about a given system to formally measure this libration. Instead we classify each pair by the weaker criterion
of proximity to the commensurability defined by the period ratio. To this end, we define the quantity $\Delta$ as
\begin{equation}
\Delta = \frac{P_i}{P_j} \frac{k}{k+1} - 1
\end{equation}
to represent the proximity of the period ratio $P_j/P_i$ to the first order commensurabilty $(k+1)/k$, for integer k..  For a pair that is in resonance, $\Delta$
will be close to zero, although there will be a small amplitude oscillation about zero during the libration of the resonant angle. For planetary
masses in the range 1--10 $M_{\oplus}$ and eccentricities $\sim 0.001-0.1$, resonant libration widths are expected to yield resonant
boundaries $0.001 < \left| \Delta \right| <0.01$ (based on the formulae in \cite{MD2000}.)

As the system undergoes instability, planets will scatter and collide\footnote{For planets close to the star, the dominant mode of planet loss is collision
with another planet, although a small fraction may be scattered into the central star. Ejection from the system, while common for dynamical instability 
in the outer parts of planetary systems, is negligible in this case.}, and the period ratios will change. The consequence of this evolution
is that fewer pairs will be in mean motion resonances, and we expect $\Delta$ to be larger for remnant pairs in which one, or both, of the
members have undergone a scattering and collision.
We wish to determine the likely threshold between the two regimes, as this will enable
us to define which planets can be regarded as pristine -- unaffected by late collisions -- and which could be sculpted by giant impacts.

\subsection{Collision boundaries in dynamical architectures}

The standard model for the dynamical dissolution of primordial planetary systems is that they form compact,
resonant chains during the protoplanetary disk phase, when dissipation from the gaseous disk suppresses the growth
of eccentricities, and so prevents orbits from crossing those of neighbour planets. Once the gas disk dissipates, the
mutual perturbations of the planets excite the eccentricities until the orbits cross, resulting in close planetary passages,  scattering,
 and orbital  instability. 
This results in, mostly, planetary
 collisions until  a final, non-resonant, more widely spaced configuration is achieved. The resulting planetary system will
 contain planets that have experienced giant impacts with other planets, although some may possibly avoid such fates.
 We wish to identify final architectures in which these different classes of planets are to be found.
 
 In a fully evolutionary calculation, the path to instability requires a slow, chaotic evolution of the resonant angles in a multiplanet system
which eventually leads to dynamical instability. However, the exact path to instability is not our concern here -- we are primarily interested in the
outcomes of the instability. The orbit-crossing and planetary scattering that characterises this process will also erase the memory of how the
system actually approached instability. Thus, we can accelerate our calculation by randomizing the resonant angles at the outset, which accelerates
the system along the path to instability.

In this regard we follow an approach similar in spirit to \cite{GB23} or \cite{LCCD25}. We simulate a set of planetary systems orbiting a $1 M_{\odot}$ star. We distribute a fixed amount
of planetary mass between $N$ planets, where $N$ is chosen randomly, with uniform probability, between 5--8 (inclusive). The masses of the planets are
also chosen randomly, with the restriction that no single planet contains over 50\% of the total mass. The largest planet was also chosen to be one of the three
innermost planets, and set at 0.06~AU. This was to reflect the fact that Type~I migration is faster for more massive planets and so a more massive planet would
migrate to the disk inner edge and force interior planets into the gap. The restriction that one of the three innermost reside at the inner edge was to prevent interior
planets being forced too far inwards, to physically unrealistic proximity to the star.

Subject to the above restrictions, the initial planetary semi-major axes were chosen to lie on first order commensurabilities $(q+1)/q$, where $q$ is
distributed with equal probability between $q=1$ and $q=5$.
 As noted above, we wish
to examine the consequences of the dynamical instability that occurs once the dampening effects of the gas disk are removed, so other orbital elements were chosen to promote moderate instability. Eccentricity was chosen from a one-sided
normal distribution with mean 0 and dispersion 0.05. The argument of periapsis was set at 0 for all, while the longitude of the ascending node and the mean
anomaly were chosen randomly over $2 \pi$. Inclinations were uniformly randomised between 0 and $0.5^{\circ}$. 

The planetary radii were calculated from the masses, assuming a mean density of 1.6$g/cm^3$ (roughly that of Neptune). We performed simulations with 
a range of total masses -- 250 runs with a total mass $9.75 M_{\oplus}$, 300 runs with total mass $15 M_{\oplus}$ and 50 runs with total mass $30 M_{\oplus}$.
Numerical simulations of the dynamical evolution were performed with the IAS15 integrator in the REBOUND package \citep{RS15}. Collisions were resolved
as perfect mergers in which both mass and composition were preserved. Integration times were for 2~Myr, which proved sufficient to resolve the dynamical
evolution of the instability, once triggered. Final outcomes were classified as follows.

Neighbouring planet-pair period ratios of surviving planets were averaged over the final 10 kyr of each simulation, and compared against the same discrete set of first-order commensurabilities from $q=1$ to $q=5$.
The smallest offset from any of these was recorded as that pair's $\Delta$ value. In order to assure that the measured $\Delta$
reflected a final stable configuration, each run was 
 was also screened for the possibility that the run ended during a period of instability. If 
 any planet's semi-major axis varied by more than 0.075 AU over the final 10 kyr window, that final configuration was not included in our analysis. Ultimately,
 none of the simulated systems failed this test.
 


\begin{figure}
\centering
\includegraphics[width=1.0\linewidth]{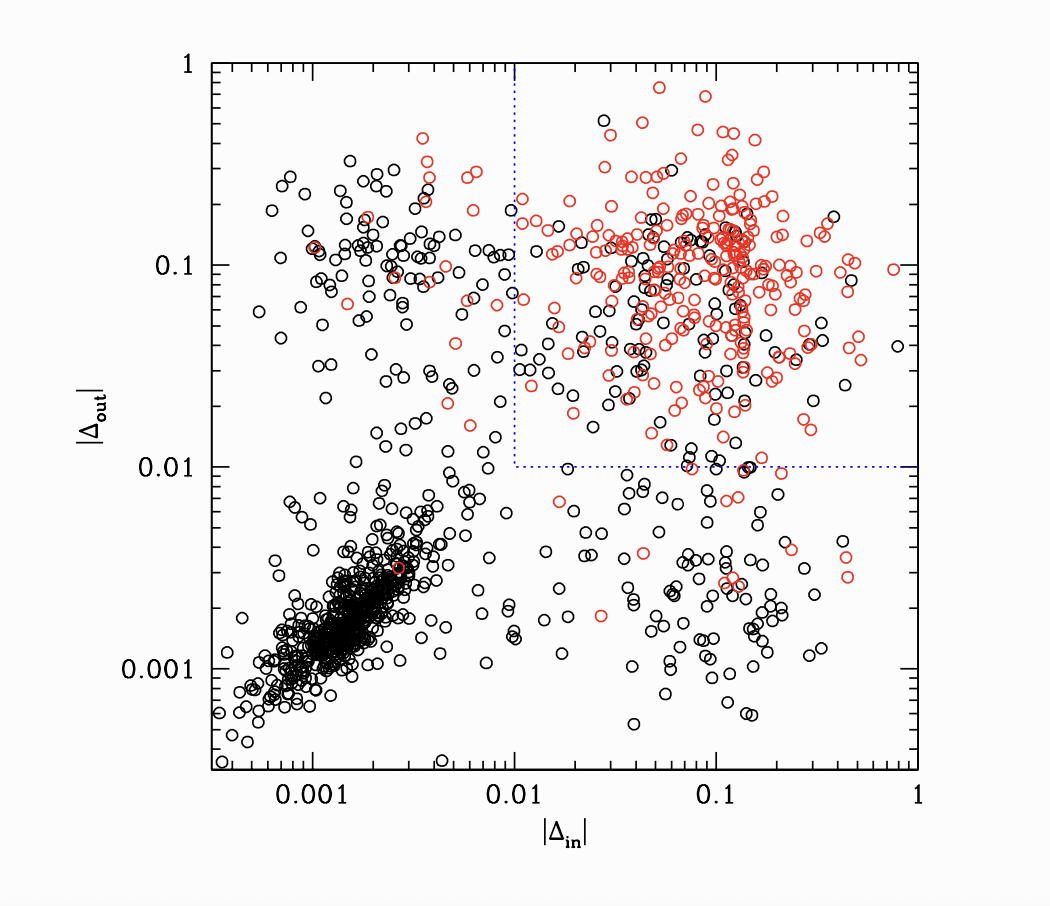}
\caption{The  black open points represent planets that do not experience a collision during the dynamical evolution of the resonant chain. The
open red points indicate planets that did experience a collision, resulting from instabilities that occur during the dynamical evolution. The blue dotted lines delineate 
the approximate transition  ($\left| \Delta \right|=0.01$) between regimes where the majority has/has not experienced a collision.
 The odds of a given planet having experienced a giant collision
during the dynamical instability of the system are increased if both pairs lie above this threshold. \label{DD7}}
\end{figure}

Figure~\ref{DD7} shows the resulting distribution of surviving planets at the end of the dynamical simulations. We mark as solid black points those planets
that did not undergo a collision during the evolution, and the solid  red points are those planets which did experience a collision. We plot only planets
with both an inner and outer neighbour at the end of the simulation. 
  This proximity to resonance is represented by
 $\Delta$ with respect
to the planet interior to it ($\Delta_{in}$) and the planet exterior to it ($\Delta_{out}$). In each case, the value is calculated relative to the
closest first order resonance. 

It is clear that systems with low $\left| \Delta_{in} \right|$ and $\left| \Delta_{out} \right|$ have a high probablity to have passed
through their evolution unscathed -- without experiencing any giant impacts. Similarly, planets with large values in both quantities
are likely to have undergone at least one collision. As a rough division between these two regimes, we note that $\left| \Delta \right| \sim 0.01$ divides
the parts of the parameter space dominated by collisions from those dominated by unaffected survivors. Planets with both inner and outer $\left| \Delta \right|$ below
this threshold are overwhelmingly pristine, while those with both above this threshold have a high probability of experiencing at least one collision. Those
parts of parameter space where only one of the criteria is satisfied are more ambiguous.

\begin{figure}
\centering
\includegraphics[width=1.0\linewidth]{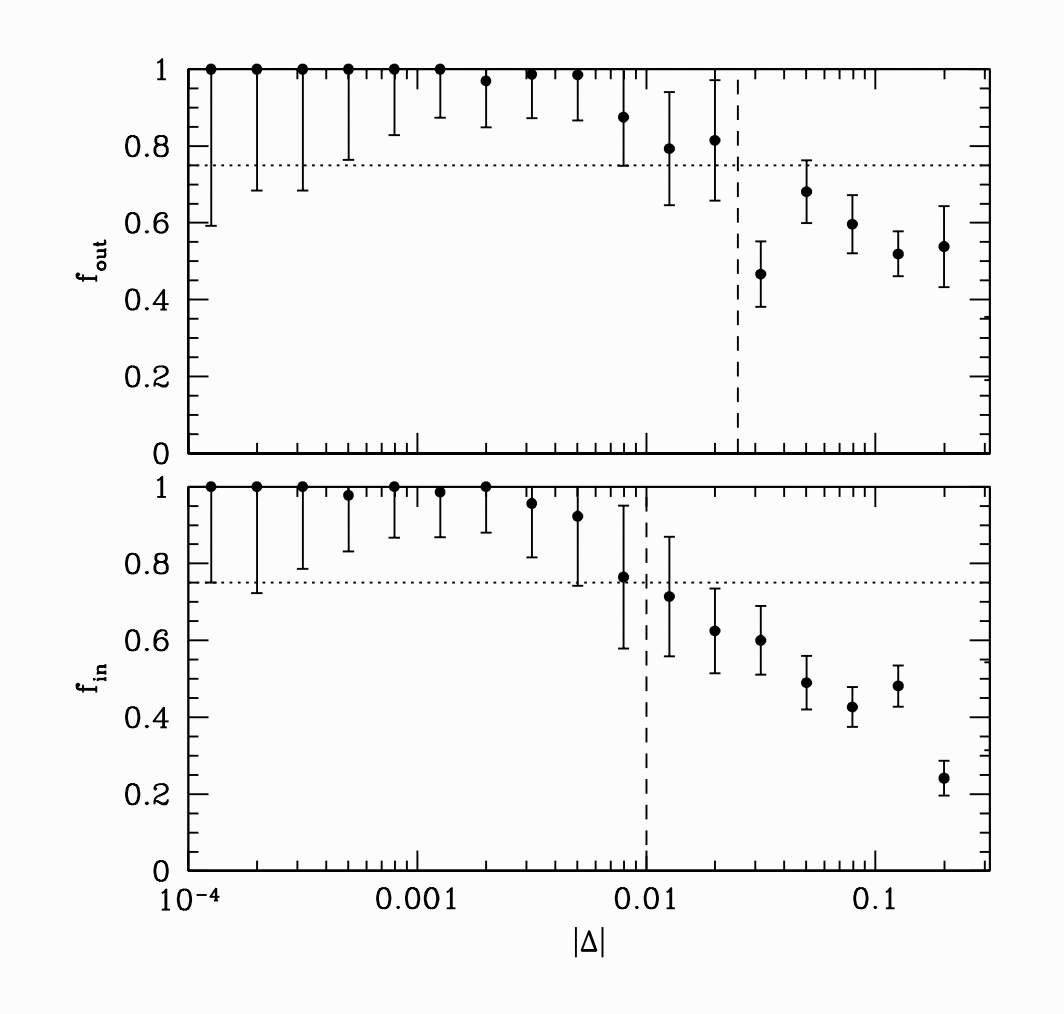}
\caption{The solid points in the upper panel show the fraction of planets in each bin of $\left| \Delta \right|$ that managed to
avoid a collision during the dynamical instability of compact planetary systems. The error bars indicated Poisson counting errors. This
panel represents the surviving outer planets in each chain. The lower panel shows the equivalent fraction for the surviving inner planets
of these chains. The horizontal dotted line indicates a survival fraction of 75\%. The dashed lines indicate the values at which
each curve crosses the horizontal dotted line. \label{Dinout}}
\end{figure}

Focussing on the planets with both interior and exterior neighbours does not make use of the planets on either end of the
observed planetary chain. With only one $\Delta$ to characterise the dynamical state, these planets are less well divided 
into collisional categories. Figure~\ref{Dinout} shows the fraction ($f_{in}$) of inner -- in the lower panel -- and outer  ($f_{out}$) -- the 
lower panel -- planets that avoided having a collision, binned as a function of the $\left| \Delta \right|$ with respect to the nearest neighbour
(so $\Delta_{out}$ applies for the inner planet of a chain, and $\Delta_{in}$ for the outer planet of a chain).

We define a threshold of 75\% to classify our pristine fraction in both cases, which amounts to a criterion $\left| \Delta_{out} \right| <0.01$ for inner
planets, and $\left| \Delta_{in} \right|<0.025$ for outer planets.

\section{Dynamically Selected Mass-Radius relations}
\label{DynMR}

The attraction of a criterion based on $\Delta$ is that this information is easily calculated for any planet pair with measured orbital periods. We have
examined the data from the NASA Exoplanet Archive (as of May 1 2026), to select a set of planets according to the above criteria. We select only planets
that have both an inner and an outer companion, and mass and radius both measured to better than 50\% accuracy. We also restrict our sample 
to masses $<100 M_{\oplus}$ to avoid  overlap with giant planets -- which are expected to follow different evolutionary pathways. Even this criterion
may be too generous -- we will return to that discussion later.
 Values of $\Delta$ were measured
relative to all first order commensurabilities $(q+1)/q$ from $q=$1--5, and the one with the minimum absolute value was adopted. We now seek to
classify each planet according to the likelihood that it has experienced a late stage collision, as characterised by the above simulations.

To obtain a systematic division into different classes, we used a machine learning algorithm. 
 This dataset was used to train a k-nearest neighbours (K-NN) model with $k \approx 0.5\sqrt{N}$, corresponding to k = 25 from a training set of N = 2665. Contributions of a point to its respective class was weighted by a Gaussian function of the distance, in the logarithmic $\left| \Delta \right|$ space, to the query point. The model was then deployed on the observed planet set with both values of   $\left| \Delta \right|$, and probabilistic predictions were generated for each planet based on their nearest neighbours’ weights. Planets were classified as collision products or pristine if they were assigned to that class with a probability $> 75$\% and classified as `unassigned' if they could not be categorised at that level of confidence. To classify the full range of observed pairs did require extrapolating beyond the range of final separations in the simulations.
 To flag potentially unreliable extrapolations, predictions were classified as outliers if either $\left| \Delta \right|>1$ -- i.e. orbital period ratios $>4$.  

\begin{figure}
\centering
\includegraphics[width=1.0\linewidth]{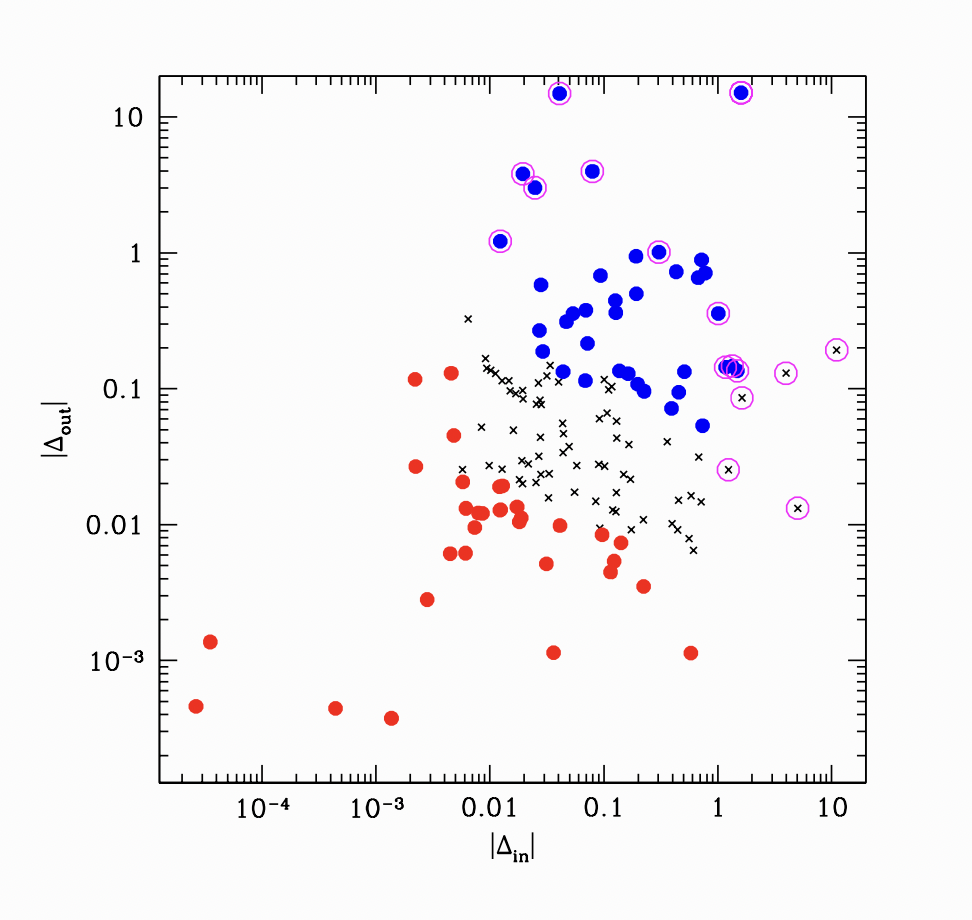}
\caption{The  red solid points represent observed planets with probability $p>0.75$ of having survived without undergoing a collision.
The blue solid circles represent observed systems with a probability $p>0.75$ of having undergone a collision in their evolutionary
history. The crosses represent systems that cannot be classified securely into either camp.  We also highlight, with magenta circles, those
systems which lie outside the training set and whose description within the model is uncertain.\label{DDp}}
\end{figure}

Following the above procedure, we classify each planet as either a collision product, pristine, or unassigned, according to its
position in Figure~\ref{DD7}, which we then reproduce in Figure~\ref{DDp}, showing each planetary classification.
We show likely pristine planets as filled red circles, and likely collision products in blue. Unassigned planets are shown as black crosses.
In addition, we have highlighted those systems (with magenta circles) which were flagged as falling outside the training set, according to the above criteria.
 The presence of such pairs in the observed data may reflect the fact that, by including planets up
to $100 M_{\oplus}$, we may be including planets that form via a different process and are therefore not well represented by the
conditions of our simulations. Alternatively, most of our observed systems are detected in transit, and not every planet in a system
is guaranteed to transit. So, we expect some of the observed pairs in our sample to have anomalously large separations simply
because there is a planet in between that has been missed. Thus, we include these planets in our discussion but note their
potentially anomalous nature.

In constructing a mass--radius relation for different planet classes, we  must also account for the fact that the 
 measurements of mass for these low mass systems is drawn from a heterogenous set of methods. Our preferred source is measurements of
the radial velocity amplitude, but many planets have mass determinations by measuring transit timing variations (TTV).  
There is a history of moderate disagreements between TTV mass measurements  and radial velocity measurements of the same systems \citep{WL13,Steff16,HL17,MM17,Lel23}. 
\cite{Lel23} showed that TTV measurements at low signal-to-noise can be biased, and used the corrected sample in the study of planetary mass
and density difference as a function of proximity to resonance \citep{LDB24}. As a consequence,
 we are selective about which TTV mass measurements we include. 
We do not use the results from some of the classic early studies \citep[e.g][]{HL17}, because of the biases discussed in \cite{Luq23}, which
results in values for some systems are notably different
from more recent measurements. However, we do include TTV results from more modern analyses with updated understanding of the
systematic errors \citep{Luq23} and also include results from custom TTV studies of individual systems anchored with dynamical simulations
and radial velocity data \cite[e.g.][]{JHL14,Mills16,Agol21}

\subsection{Pristine Planets}

\begin{figure}
\centering
\includegraphics[width=1.0\linewidth]{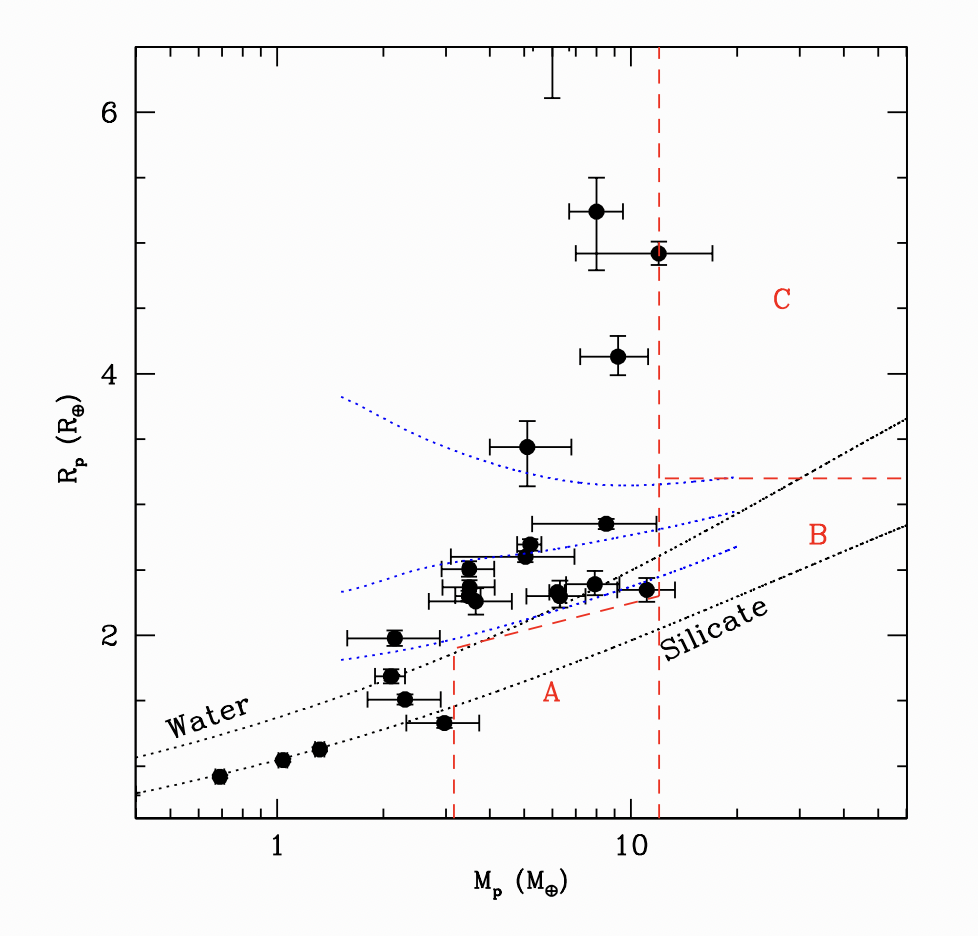}
\caption{The solid points indicate the mass-radius relations for planets that are still bound in a chain of first order mean motion resonances, and
thus unlikely to have experienced a collision since the dissipation of the gas disk. The black dotted lines are the mass-radius relation for
silicate planets and water planets, from \cite{SKH07}. 
 The three blue dotted lines represent a rocky planet with a Hydrogen envelope of 2\% by mass, from \cite{LF14}.
The upper one is for an age of 0.1~Gyr and 1000 times the Solar irradiation. The middle curve is for 10~Gyr and the same irradiation.
The lowest curve is for 10~Gyr but 0.1 Solar irradiation. These three curves illustrate the degree to which the astrophysical context can
affect radius.
 The red dashed lines indicate three regions (labelled A, B and C) that lack planets,
even though planets do exist in these regions in the wider planetary database.\label{MR1}}
\end{figure}

Figure~\ref{MR1} shows the mass--radius information for planets which are classified as pristine by the above procedure (i.e. the red points in
Figure~\ref{DDp}).
 These
are the planets most likely to have not undergone any significant collisional history at late times and so should best represent the primordial planetary composition, in
the sense that they should have the same structure as they did when they were locked in the resonant chain at the end of planetary migration.
 The sample is small (27, including four from the Trappist-1 system), by virtue of our rather strict criteria
on both dynamical grounds and regarding the selection of measured masses. The planets included are listed in Table~\ref{PristineMR}.
  An advantage of being strict, however, is that the data show a clear pattern, with a division into two classes. There is a clear
  rocky planet branch, which follows the mass-radius relation for a Silicate equation of state, and contains planets only with
  $M < 3 M_{\oplus}$. A second population is composed of higher mass ($M > 2 M_{\oplus}$) bodies with a non-negligible
  radius contribution from lower density material. The lower envelope of this group follows the water equation of state, but many
  have radii which suggest Hydrogen envelopes. These can be explained by a Hydrogen mass fraction $\sim 2\%$, if they sample
  a range of ages and irradiances. The planets in this population also all have masses $<12 M_{\oplus}$. This is consistent with
  the core accretion model \citep{DJS82,PHD96}, in which more massive planetary cores are able to capture a significant mass of gas from the nebula
  and grow to giant planet sizes.

\begin{table}[ht]
\caption{Planets with Low Collision Probability \label{PristineMR}}
\begin{tabular}{|lllll|}
\hline
$M_p$  & $R_p$  & Period  & Name & Ref. \\
 ($M_{\oplus}$) & ($R_{\oplus}$)  & (days) &   &\\
\hline
$0.39^{+0.01}_{-0.01}$ & 0.79$^{+0.01}_{-0.01}$ & 4.049 & Trappist-1~d & 1 \\
$0.69^{+0.02}_{-0.02}$ & 0.92$^{+0.01}_{-0.01}$ & 6.6101 & Trappist-1~e & 1 \\
$1.04^{+0.03}_{-0.03}$ & 1.05$^{+0.01}_{-0.01}$ & 9.2075 & Trappist-1~f & 1 \\
$1.32^{+0.04}_{-0.04}$ & 1.13$^{+0.02}_{-0.01} $& 12.352 & Trappist-1~g & 1\\
$2.10^{+0.20}_{-0.21}$ & 1.69$^{+0.05}_{-0.05}$ & 12.070 & Kepler-54~c & 2\\
$2.15^{+0.73}_{-0.57}$ & 1.98$^{+0.06}_{-0.06}$ & 12.51 & Kepler-85~c & 2 \\
$2.30^{+0.60}_{-0.50}$ & 1.51$^{+0.04}_{-0.04} $ & 13.78 & Kepler-138~c & 3 \\
$2.97^{+0.76}_{-0.65}$ & 1.33$^{+0.04}_{-0.04}$ & 4.6447 & Kepler-80~e & 4 \\
$3.48^{+0.29}_{-0.29}$ & 2.30$^{+0.04}_{-0.04}$ & 9.964 & TOI-178~e & 5 \\
$3.50^{+0.63}_{-0.57}$ & 2.37$^{+0.05}_{-0.05}$ & 7.0534 & Kepler-80~b & 4 \\
$3.65^{+0.97}_{-0.97}$ & 2.26$^{+0.10}_{-0.10}$ & 4.695 & TOI-663~c & 6 \\
$3.84^{+0.39}_{-0.40}$ & 2.05$^{+0.07}_{-0.07}$ & 8.9187 & Kepler-60~c & 2 \\
$5.04^{+1.89}_{-1.94}$ & 2.60$^{+0.04}_{-0.04}$ & 41.0585 & HD~110067~f & 7\\
$5.10^{+1.70}_{-1.20}$ & 3.44$^{+0.20}_{-0.30}$ & 9.8456 & Kepler-223~c & 8 \\
$5.20^{+0.39}_{-0.42}$ & 2.70$^{+0.04}_{-0.05}$ & 6.5576 & TOI-178~d& 5 \\
$6.00^{+0.70}_{-0.70}$ & 6.53$^{+0.42}_{-0.42}$ & 12.4014 & V1298~Tau~d & 9 \\
$6.10^{+1.30}_{-1.30}$ & 3.18$^{+0.09}_{-0.09}$ & 28.51 & HD191939~c & 10 \\
$6.20^{+0.31}_{-0.31}$ & 2.33$^{+0.01}_{-0.01}$ & 5.6605 & TOI-270~c & 11 \\
$6.31^{+1.13}_{-1.23}$ & 2.30$^{+0.12}_{-0.09}$  & 3.5600 & K2-138~c & 12\\
$6.90^{+1.10}_{-1.10}$ & 9.32$^{+0.18}_{-0.18}$ & 130.1858 & Kepler-51~d & 13 \\
$7.92^{+1.38}_{-1.35}$ & 2.39$^{+0.10}_{-0.08}$ & 5.405 & K2-138~d & 12 \\ 
$8.00^{+1.50}_{-1.30}$ & 5.24$^{+0.26}_{-0.45}$ & 14.7887 & Kepler-223~d & 8\\
$8.57^{+3.31}_{-3.25}$ & 2.85$^{+0.04}_{-0.04}$ & 20.5196 & HD~110067~d & 7  \\
$9.20^{+2.00}_{-2.00}$ & 4.13$^{+0.16}_{-0.14}$ & 55.997 & HD~28109~c & 14 \\
$11.1^{+2.25}_{-1.93}$ & 2.35$^{+0.09}_{-0.09}$ & 8.140 & Kepler-24~b& 2 \\
$12.0^{+5.0}_{-5.0}$ & 4.92$^{+0.09}_{-0.09}$ & 369.0 & HIP~41378~e & 15 \\
$32.4^{+1.70}_{-1.70}$ & 7.00$^{+0.20}_{-0.20}$ & 7.922 & K2-19~b& 15\\
\hline
\end{tabular}

{{\bf References:} [1] \cite{Agol21};
   [2] \cite{Lel23}; 
 [3] \cite{PBA23} ; [4] \cite{Mac21};
 [5] \cite{LDD24}; 
   [6] \cite{CBA24};   [7] \cite{Luq23}; [8] \cite{Mills16};
  [9] \cite{LPD26}; 
  [10] \cite{PLB24}; 
  [11]  \cite{KVG22};[12] \cite{Lopez19}; [13] \cite{Masuda24};
   [14] \cite{BRG25};
    [15] \cite{HSB25}}
\end{table}  

In Figure~\ref{MR1} we also define three regions, labelled A, B and C.
Region~A would comprise rocky planets more massive than 3$M_{\oplus}$. As we will see below, there
are planets that can be found with values in this region, but their absence here suggests their prescence in Region~A 
may not be primordial. It appears that purely rocky planets are limited to masses $< 3 M_{\oplus}$
(although the sample is admittedly small).
Regions~B and C are also populated in diagrams to follow. These represent 
the domain of massive planets ($> 12 M_{\oplus}$). We divide this region further into those that remain relatively compact (radius $< 3.2 R_{\oplus}$) -- region B -- and those that have larger Hydrogen envelopes -- region C. We will also discuss these further, as we cover more of the planet sample.

\subsection{Collision Products}

\begin{figure}
\centering
\includegraphics[width=1.0\linewidth]{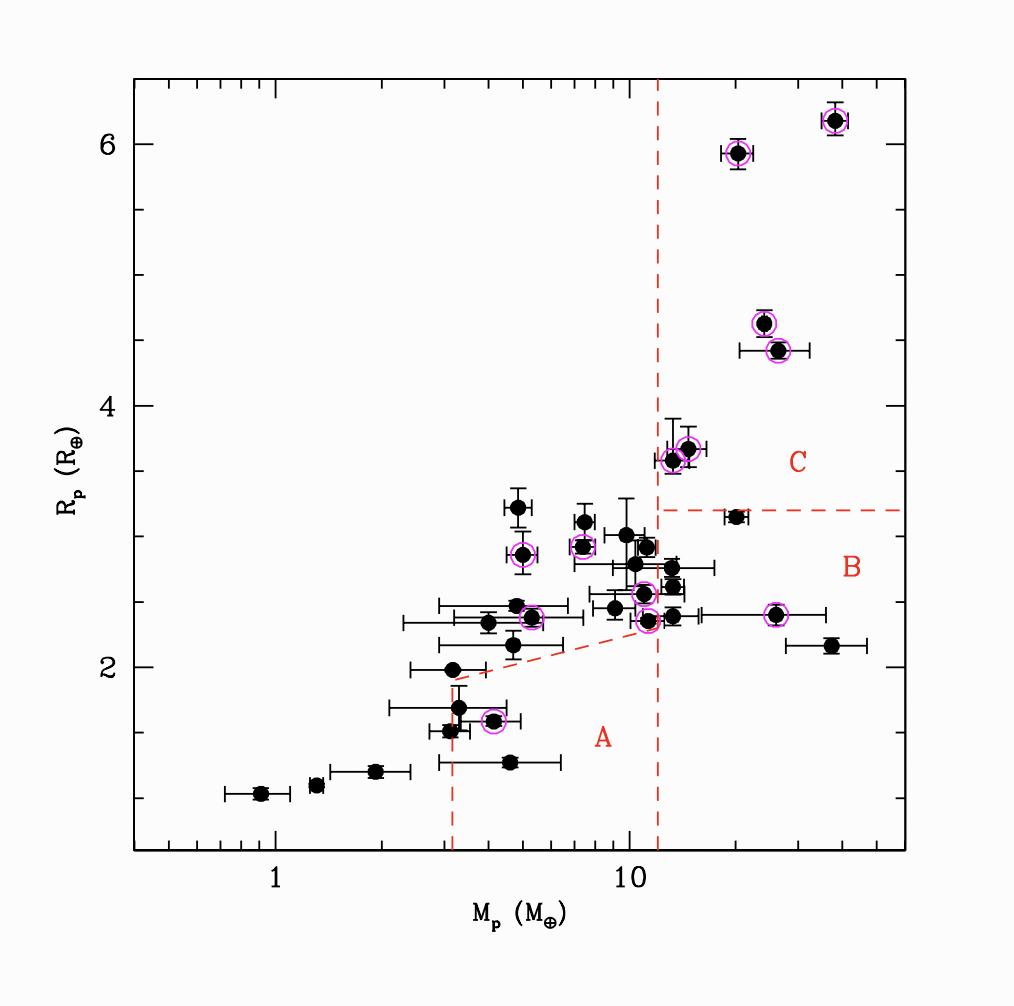}
\caption{The solid black points represent the sample defined by the blue points in Figure~\ref{DDp} -- those that have
a $>75\%$ probability of having undergone a collision. The magenta circles indicate planets whose classification is uncertain because the
observed $\left| \Delta \right|$ lay outside of the regime of the training data.
The regions A, B and C are the same as in Figure~\ref{MR1}. \label{MR2}}
\end{figure}

Although the properties of pristine planets are interesting from the point of view of understanding how planets form, the
effect of giant impacts on planetary structure carries the best information concerning the importance of planet-scale collisions in the
evolution of these planetary systems.
In Figure~\ref{DDp} we  defined a subset
of planets with a high probability of having undergone a collision -- the blue points. This sample is shown in Figure~\ref{MR2} and
enumerated in Table~\ref{CollProducts}.

\begin{table}[ht]
\caption{Planets with High Collision Probability  \label{CollProducts}}
\begin{tabular}{|lllll|}
\hline
$M_p$  & $R_p$  & Period  & Name & Ref. \\
 ($M_{\oplus}$) & ($R_{\oplus}$)  & (days) &  & \\
\hline
$0.91^{+0.19}_{-0.19}$ & 1.03$^{+0.04}_{-0.04}$ & 2.7549 & LP791-18~d & 1 \\
$1.31^{+0.06}_{-0.06}$ & 1.10$^{+0.01}_{-0.01}$ & 2.4219 & Trappist-1~c & 2 \\
$1.92^{+0.49}_{-0.49}$ & 1.20$^{+0.05}_{-0.05}$ & 3.648 & GJ9827c & 3 \\
$3.12^{+0.42}_{-0.40}$ & 1.51$^{+0.05}_{-0.05}$ & 6.765 & HD219134~c & 4 \\
$3.17^{+0.76}_{-0.76}$ & 1.98$^{+0.10}_{-0.10}$ & 18.8027 & TOI-1266~c & 5\\
$3.30^{+1.20}_{-1.20}$ & 1.69$^{+0.17}_{-0.17}$ & 24.6466 & K2-3~c & 6\\
$4.00^{+1.70}_{-1.70}$ & 2.34$^{+0.08}_{-0.08}$ & 12.82 & Kepler-100~c & 7 \\
$4.14^{+0.79}_{-0.80}$ & 1.59$^{+0.04}_{-0.04}$ & 8.1317 &  Kepler-65~d & 8 \\
$4.60^{+1.80}_{-1.70}$ & 1.27$^{+0.04}_{-0.04}$ & 7.0600 & K2-233~c & 9 \\
$4.70^{+1.80}_{-1.80}$ & 2.17$^{+0.11}_{-0.11}$ & 16.1457 &  Kepler-102~e & 7\\
$4.80^{+1.90}_{-1.90}$ & 2.47$^{+0.04}_{-0.04}$ & 7.8852 & TOI-5624~c & 10\\
$4.85^{+0.44}_{-0.42}$ & 3.22$^{+0.15}_{-0.15}$ & 12.2800 & Kepler-26~b  & 11\\
$5.00^{+0.50}_{-0.50}$ & 2.86$^{+0.18}_{-0.15}$ & 12.9278 & TOI-5789~c & 12\\
$5.30^{+2.10}_{-2.10}$ & 2.38$^{+0.07}_{-0.07}$ & 15.7711 & Kepler-139~b & 7\\
$7.39^{+0.62}_{-0.62}$ & 2.92$^{+0.05}_{-0.05}$ & 25.5027 & TOI-1203.01 &  13\\
$7.48^{+0.49}_{-0.48}$ & 3.11$^{+0.14}_{-0.14}$ & 17.2559 & Kepler-26~c & 11\\
$9.10^{+1.30}_{-1.30}$ & 2.45$^{+0.14}_{-0.09}$ & 5.9041 &  TOI-1246~c & 14 \\
$9.80^{+1.24}_{-1.30}$ & 3.01$^{+0.28}_{-0.42}$ & 29.854 & HD3167~c & 3 \\
$10.4^{+1.30}_{-1.10}$ & 2.42$^{+0.07}_{-0.07}$ & 36.527 & HD~224018~c & 15\\
$11.1^{+3.3}_{-3.3}$ & 2.56$^{+0.07}_{-0.07}$ & 9.6740 &  Kepler-48~c & 7\\
$11.2^{+0.65}_{-0.63}$ & 2.92$^{+0.07}_{-0.07}$ & 27.592 & HD136352~c & 16 \\
$11.3^{+1.24}_{-1.24}$ & 2.36$^{+0.02}_{-0.02}$ & 95.2943 & Kepler-10~c & 6 \\
$13.2^{+4.23}_{-4.23}$ & 2.76$^{+0.07}_{-0.07}$ & 7.4931 & TOI-1260~c & 17 \\
$13.3^{+1.5}_{-1.5}$ & 3.58$^{+0.32}_{-0.10}$ & 9.0310 & WASP-47~d & 3 \\
$13.3^{+2.40}_{-2.40}$ & 2.39$^{+0.07}_{-0.07}$ & 13.5208 & Kepler-106~c & 7 \\
$13.3^{+0.98}_{-0.98}$ & 2.62$^{+0.06}_{-0.06}$ & 25.7127 & TOI-561~d & 18 \\
$14.7^{+1.84}_{-1.89}$ & 3.67$^{+0.17}_{-0.14}$ & 15.624 & EPIC~249893012~c & 19 \\
$15.2^{+1.3}_{-1.6}$ & 5.22$^{+0.07}_{-0.07}$ & 12.7207 & Kepler-25~c & 8 \\
$20.1^{+1.6}_{-1.5}$ & 3.15$^{+0.04}_{-0.04}$ & 18.2616 & TOI-2141~b & 20 \\
$20.3^{+2.13}_{-2.11}$ & 5.93$^{+0.11}_{-0.12}$ & 5.4147 & TOI-4010~c & 21\\
$26.0^{+10.0}_{-10.0}$ & 2.40$^{+0.08}_{-0.08}$ & 82.2000 & Kepler-129~c & 7 \\
$26.4^{+5.9}_{-5.9}$ & 4.42$^{+0.06}_{-0.06}$ & 7.8344 & Kepler-411~c & 22\\
$37.3 ^{+9.6}_{-9.6}$ & 2.16$^{+0.06}_{-0.06}$ & 7.1079 & HD63433~b & 14 \\
$38.2^{+3.3}_{-3.3}$ & 6.18$^{+0.14}_{-0.11}$ & 14.7089 & TOI-4010~d& 21 \\
$77.0^{+8.00}_{-8.00}$ & 10.3$^{+0.24}_{-0.24}$ & 22.3430 & KOI-94~d & 7 \\
\hline
\end{tabular}

{ {\bf References:} [1] \cite{GMK25} [2] \cite{Agol21}; [3] \cite{HSB25}; 
[4] \cite{HDT25};
 [5] \cite{GMV25};  [6] \cite{BDM23};  [7] \cite{WIH24}; [8] \cite{MHW19}; 
 [9] \cite{BGA23}; 
   [10] \cite{BGLO26};   [11] \citep{Lel23}; [12] \cite{BNS26}; 
   [13] \cite{GAS25} ; [14] \cite{PLB24}; [15] \cite{DNA25}; 16] \cite{DEA21};
   [17] \cite{LCH23};  [18] \cite{PZB24}; [19] \cite{HPA20}; [20] \cite{LLC25};   
   [21] \cite{KVH23};  [22]  \cite{SIG19}
   }
\end{table}

This population of 35 planets shows the overall same division -- a lower mass population of
rocky planets and a higher mass population with a lower density envelope. However, there
are some important differences.

Region~A is no longer completely empty -- there are three planets plausibly in this
region, which may represent merged remnants of smaller planets. There is also now 
a significant number of planets with masses $> 12 M_{\oplus}$ in regions B and C.
There also appears to be a pileup of planets with mass $\sim 10 M_{\oplus}$.

The simplest explanation is that these populations are indeed generated by collisions
of smaller planets during dynamical instability. However, the distribution of radii in
Figure~\ref{MR2} suggests some caution is warranted. The planets that populate region~B
do indeed seem to be massive planets with  Hydrogen inventories  comparable to those
of the 
corresponding lower mass planets. However, the group that now occupies region~C have
considerably larger Hydrogen envelopes than the lower mass planets. These are not
plausible collision candidates because they would have had to have {\bf increased} their
Hydrogen inventories considerably during the collsion. Rather, this suggests that our sample definition may include
some planets which form by a different mechanism. This is further supported by the appearance
of  magenta
circles in Figure~\ref{MR2} -- these are systems whose separations fall outside the bounds 
of our simulations and 3/5 of the planets in group C fail this criterion. These planets, and their companions, are too widely
spaced to plausibly result from dynamical instability of more closely packed configurations. We will
discuss the origins of these planets in \S~\ref{Infer}.


\subsection{Unassigned Planets}
\label{Unknown}

\begin{figure}
\centering
\includegraphics[width=1.0\linewidth]{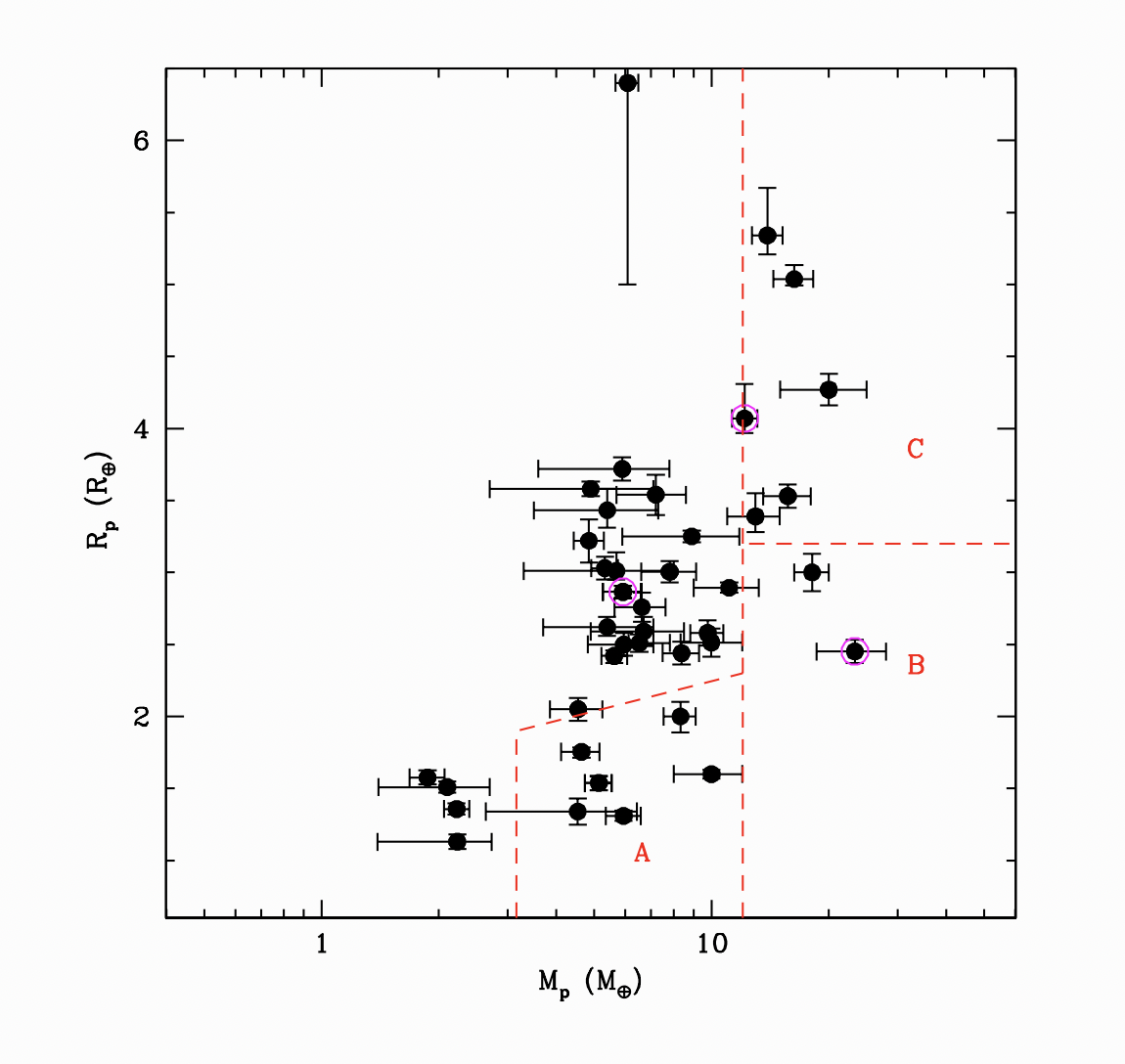}
\caption{The solid black points are those for which our model cannot confidently assign
to either the pristine or collision groups. We see that the region A is better populated than in either of the other two plots, but
region B is not. As in Figure~\ref{MR2}, magenta circles indicate systems that fall outside of the training set. \label{MR3}}
\end{figure}

Figures~\ref{MR1} and \ref{MR2} contain those planets that can be classified
as likely to either have had, or not to have had, a collision -- the blue and red
points in Figure~\ref{DDp}. However, this still does not include those indicated
as crosses in Figure~\ref{DDp} -- the planets that could not be classified, one way or
the other, with a probability $>75\%$. 
This sample is shown in Figure~\ref{MR3}.  

The distribution in this diagram contains elements of both Figures~\ref{MR1} and \ref{MR2}, as
might be expected. Once again, there are several planets in region~A, and two in region~B.
There are several planets in region~C, suggesting contamination by another
formation mechanism. On the other hand, the distribution of sub-Neptune planets shows a
mass-radius distribution similar to the corresponding group in Figure~\ref{MR1}, and not
the systematic shift to larger masses observed in Figure~\ref{MR2}.

One possible explanation for this diversity is the potential contribution from  processes, other than collisions, 
that affect $\Delta$, which would blur the lines between our different planetary groups. Our classification is quantified within the framework of dynamical
instability, wherein planets cross orbits, scatter and collide. Small $\left| \Delta \right|$ 
indicate proximity to resonance and a low probability of instability, whereas large $\left| \Delta \right|$ 
is suggestive of instability that results in collisions and mergers (the removal of one of the planets from
the chain will obviously result in much larger separations). Unassigned planets occupy a regime
of $\left| \Delta \right|$ intermediate between these extremes. This is to be expected, as some
planets will be neighbours to potentially pristine planets on one side and collision products on
the other, and therefore residing in an intermediate part of Figure~\ref{DDp}.

There are several alternative mechanisms that can also generate moderate deviations in
$\left| \Delta \right|$. Any dissipative process acting on a resonant pair can generate `resonant
repulsion' \citep{LW12,BM13}, causing the $\left| \Delta \right|$ to increase, to a degree that
will depend on the level of dissipation. This can be a
consequence of tidal damping of eccentricity or as the result of scattering of
planetesimals \citep{WML24}. Similarly, loss of atmospheric mass due to photoevaporation
from the central star \citep{OW13,LF13} can cause $\left| \Delta \right|$ to increase, if some of the
angular momentum of the evaporated material is returned to the planets by wind--magnetosphere
coupling \citep{WL23}.

 Several of the mechanisms cited above
are also strongly dependant on the distance from the host star. Thus, it motivates us to consider
the planets on the edges of the chains.

\begin{table}[ht]
\caption{Unassigned Planets.\label{MuddyMR}}
\begin{tabular}{|lllll|}
\hline
$M_p$  & $R_p$  & Period  & Name & Reference \\
 ($M_{\oplus}$) & ($R_{\oplus}$)  & (days) &  &  \\
\hline
$0.72^{+0.28}_{-0.27}$ & 0.82$^{+0.09}_{-0.07}$ & 7.908 & HD~23472~e & 1\\
$1.87^{+0.20}_{-0.19}$ & 1.58$^{+0.05}_{-0.05}$ & 7.451 & L98-59~d & 2 \\
$2.10^{+0.6}_{-0.70}$ & 1.51$^{+0.04}_{-0.04}$ & 23.0923& Kepler-138~d & 3 \\
$2.22^{+0.17}_{-0.16}$ & 1.36$^{+0.04}_{-0.04}$ & 3.691 & L98-59~c & 2 \\
$2.23^{+0.50}_{-0.84}$ & 1.13$^{+0.05}_{-0.05}$ & 34.169 & KOI-1831~d &4 \\
$3.49^{+0.63}_{-0.57}$ & 2.51$^{+0.06}_{-0.06}$ & 9.520 & Kepler-80~c & 5\\
$4.54^{+1.90}_{-1.90}$ & 1.34$^{+0.09}_{-0.09}$ & 10.312 &  Kepler-102~d & 6 \\
$4.55^{+0.70}_{-0.70}$ & 2.05$^{+0.08}_{-0.08}$ & 6.220 & LHS~1903~c & 7 \\
$4.64^{+0.53}_{-0.53}$ & 1.75$^{+0.03}_{-0.03}$ & 3.239 & TOI-178~c & 8\\
$4.85^{+0.44}_{-0.42}$ & 3.22$^{+0.15}_{-0.15}$ & 12.2800 & Kepler-26~b  & 9\\
$4.90^{+2.20}_{-2.20}$ & 3.58$^{+0.05}_{-0.05}$ & 13.7315 & TOI-5624~d & 10\\
$5.14^{+0.41}_{-0.41}$ & 1.54$^{+0.05}_{-0.05}$ & 6.4296 &  HIP29442~d & 11 \\
$5.33^{+0.43}_{-0.42}$ & 3.03$^{+0.08}_{-0.08}$ & 66.028 &  Kepler-289~d & 12 \\
$5.40^{+1.70}_{-1.70} $ & 2.62$^{+0.07}_{-0.06}$ & 5.8600 &  Kepler-65~c & 13\\
$5.40^{+1.90}_{-1.90}$ & 3.43$^{+0.15}_{-0.12}$ & 18.655 & TOI-1246~d & 14 \\
$5.63^{+0.45}_{-0.45}$ & 2.42$^{+0.04}_{-0.05}$ & 15.233 &  TOI-178~f & 8 \\
$5.70^{+2.30}_{-2.30}$ & 3.01$^{+0.13}_{-0.06}$ & 20.661 &  K2-32~c & 15 \\
$5.90^{+1.90}_{-2.30}$ & 3.72$^{+0.08}_{-0.08}$ & 27.4029 & Kepler-79~c & 16 \\
$5.93^{+0.67}_{-0.67}$ & 2.87$^{+0.04}_{-0.04}$ & 10.7784 & TOI-561~c & 17 \\
$5.95^{+0.63}_{-0.60}$ & 1.31$^{+0.04}_{-0.03}$ & 3.0730 & Kepler-80~d & 5 \\
$5.96^{+1.14}_{-1.14}$ & 2.50$^{+0.08}_{-0.08}$ & 12.566 & LHS~1903~d & 7 \\
$6.00^{+2.10}_{-1.60}$ & 7.16$^{+0.13}_{-0.13}$ & 52.090 &  Kepler-79~d & 16\\
$6.09^{+0.41}_{-0.41}$ & 6.40$^{+1.40}_{-1.40}$ & 85.315 & Kepler-51~c&18 \\
$6.53^{+1.30}_{-0.40}$ & 2.51$^{+0.06}_{-0.06}$ & 31.708 & HIP~41378~c & 15 \\
$6.63^{+0.99}_{-0.99}$ & 2.76$^{+0.10}_{-0.10}$ & 9.1506 & TOI-125~c & 19\\
$6.70^{+1.80}_{-1.80}$ & 2.59$^{+0.10}_{-0.10}$ & 10.355 & TOI-2076~b & 20 \\
$7.20^{+1.40}_{-1.40}$ & 3.54$^{+0.14}_{-0.14}$ & 21.014 & TOI-2076~c & 20 \\
$7.81^{+1.32}_{-1.20}$ & 3.01$^{+0.07}_{-0.07}$ & 10.74 & Kepler-23~c & 6\\
$8.32^{+0.78}_{-0.79}$ & 2.00$^{+0.11}_{-0.10}$ & 17.6671 &  HD23472~b & 1 \\ 
$8.38^{+0.92}_{-0.89}$ & 2.44$^{+0.08}_{-0.08}$ & 10.91 & Kepler-49~c & 9\\
$8.90^{+2.90}_{-3.00}$ & 3.25$^{+0.04}_{-0.04}$ & 21.4899 & TOI-5624~e & 10 \\
$9.77^{+0.96}_{-0.95}$ & 2.58$^{+0.09}_{-0.09}$ & 7.20 & Kepler-49~b & 9\\
$9.99^{+2.01}_{-1.66}$ & 2.51$^{+0.10}_{-0.10}$ & 12.330 & Kepler-24~c & 6 \\
$10.0^{+2.00}_{-2.00}$ & 1.60$^{+0.03}_{-0.03}$ & 4.9015 &  Kepler-107~c & 21\\
$12.2^{+0.96}_{-0.87}$  & 4.07$^{+0.24}_{-0.10}$ & 26.44 &  Kepler-82~b & 22\\
$13.1^{+5.30}_{-5.30}$ & 9.41$^{+0.57}_{-0.57}$ & 24.140 &  V1298~Tau~b &  23\\
$13.6^{+2.30}_{-2.30}$ & 2.88$^{+0.13}_{-0.13}$ & 10.8541 & Kepler-20~c & 21 \\
$13.9^{+1.30}_{-1.20}$ & 5.34$^{+0.33}_{-0.13}$ & 51.540 &  Kepler-82~c & 22\\
$13.0^{+1.98}_{-1.98}$ & 3.39$^{+0.16}_{-0.11}$ & 8.2615 &  K2-138~e & 24 \\
$15.7^{+2.28}_{-2.13}$ & 3.53$^{+0.08}_{-0.08} $ & 7.1381 &  K2-285~c & 25\\ 
$16.3^{+1.9}_{-1.9}$ & 5.04$^{+0.10}_{-0.05}$ & 8.9920 & K2-32~b & 15 \\
$18.1^{+1.90}_{-1.80}$ & 3.00$^{+0.13}_{-0.13}$ & 17.307 & K2-136~c & 26\\
$19.7^{+4.90}_{-4.90}$ & 4.27$^{+0.11}_{-0.11}$ & 7.64 &  Kepler-18~c&6 \\
$23.0^{+5.00}_{-5.00}$ & 2.45$^{+0.08}_{-0.08}$ & 157.065 & Kepler-139~c & 6\\
$43.4^{+1.60}_{-2.00}$ & 8.29$^{+0.53}_{-0.43}$ & 19.3 & Kepler-9b & 27\\
\hline
\hline
\end{tabular}
{{\bf References:} [1]\cite{BDA22};
[2] \cite{SDH26};  [3] \cite{PBA23}; 
 [4]  \cite{JHW21}; [5]  \cite{Mac21};
 [6]  \cite{WIH24}; [7] \cite{WSCC26}; 
 [8]  \cite{LDD24}; [9] \cite{Lel23}; [10] \cite{BGLO26}; 
 [11] \cite{EOK24}; [12] \cite{GKV23}; [13]  \cite{MHW19}; 
 [14]  \cite{PLB24}; [15]  \cite{HSB25};  [16] \cite{JHL14}; 
 [17] \cite{PZB24}; [18]  \cite{Masuda24}; [19] \cite{NGA20};
  [20] \cite{WDL26}; [21] \cite{BDM23}; [22] \cite{FvO19};
 [23] \cite{LPD26};  [24] \cite{Lopez19}; [25] \cite{PNL19};
 [26] \cite{MDV23}; [27] \cite{BMP19}}
\end{table}

\subsection{Inner and Outer Planets}

Our dynamical classification of planets has relied, up to now, on using the $\Delta$ of two
neighbouring pairs -- with planets both interior and exterior to the planet in question. Obviously,
this excludes the innermost and outermost planets of each chain. As shown in Figure~\ref{Dinout}
we still have some information regarding the likely histories of these planets, albeit with lower
confidence.

\begin{table}[ht]
\caption{Innermost planets of chains \label{InnerMR}}
\begin{tabular}{|lllll|}
\hline
$M_p$  & $R_p$  & Period  & Name & Ref. \\
 ($M_{\oplus}$) & ($R_{\oplus}$)  & (days) &   &\\
\hline
\multicolumn{5}{c}{Inner Planets with $\left| \Delta_{out} <0.01 \right|$} \\
\hline
$0.07^{+0.02}_{-0.02}$ & 0.64$^{+0.02}_{-0.02}$  & 10.3134 & Kepler-138~b & 1\\
$0.55^{+0.21}_{-0.21}$ & 0.75$^{+0.07}_{-0.07}$ & 3.9766 & HD23472~d & 2 \\
$1.84^{+0.60}_{-0.47}$ & 1.78$^{0.05}_{-0.05}$ & 8.3054 & Kepler-85~b & 3 \\
$2.56^{+0.43}_{-0.40}$ & 1.64$^{+0.02}_{-0.02}$ & 7.1070 & Kepler-23~b & 3 \\
$3.09^{+0.30}_{-0.30}$ & 1.86$^{+0.06}_{-0.06}$ & 8.0109 & Kepler-54~b & 3 \\
$3.10^{+1.05}_{-1.05}$ & 1.51$^{+0.11}_{-0.08}$ & 2.3531 & K2-138~b & 4 \\
$4.70^{+0.60}_{-0.60}$ & 5.08$^{+0.37}_{-0.37}$ & 8.2491 & V1298~Tau~b & 5 \\
$5.26^{+0.45}_{-0.45}$ & 1.89$^{+0.06}_{-0.06}$ & 7.1334 & Kepler-60~b & 3 \\
$5.69^{+1.78}_{-1.82}$ & 2.20$^{+0.03}_{-0.03}$  & 9.1137 & HD110067~b & 6\\
$7.40^{+1.30}_{-1.10}$ & 2.99$^{+0.18}_{-0.27}$ & 7.3845 & Kepler-223~b & 7\\
\hline
\multicolumn{5}{c}{Inner Planets with $\left| \Delta_{out} >0.01 \right|$} \\
\hline
0.32$^{+0.12}_{-0.12}$ & 0.88$^{+0.03}_{-0.03}$ & 2.25311 & L98-59~b & 8 \\
$1.37^{+0.07}_{-0.07}$ & 1.12$^{+0.01}_{-0.01}$ & 1.51083 & Trappist-1~b & 9 \\ 
$1.48^{+0.18}_{-0.18}$ & 1.28$^{+0.05}_{-0.05}$ & 3.3599 & TOI-270~b & 10 \\
$2.02^{+0.23}_{-0.23}$ & 1.40$^{+0.03}_{-0.03}$ & 0.4466 & TOI-561~b & 11 \\
$3.24^{+0.32}_{-0.32}$ & 1.47$^{+0.03}_{-0.03}$ & 0.8375 & Kepler-10~b & 12 \\
$3.28^{+0.42}_{-0.42}$ & 1.38$^{+0.05}_{-0.05}$ & 2.156 & LHS~1903~b & 13 \\
$3.51^{+0.33}_{-0.32}$ & 1.52$^{+0.05}_{-0.05}$ & 4.157 & TOI-1203~b & 14 \\
$3.73^{+0.34}_{-0.33}$ & 1.60$^{+0.06}_{-0.06}$ & 3.0929 & HD219134~b & 15 \\
$3.80^{+1.80}_{-1.70}$ & 1.54$^{+0.03}_{-0.03}$ & 3.180 & Kepler-107~b & 12 \\
$4.01^{+0.47}_{-0.47}$ & 1.34$^{+0.12}_{-0.12}$ & 6.887 & Kepler-100~b & 16 \\
$4.45^{+0.63}_{-0.66}$ & 2.27$^{+0.10}_{-0.09}$ & 2.5989 & TOI-663~b & 17\\
$4.50^{+0.32}_{-0.32}$ & 1.55$^{+0.05}_{-0.05}$ & 3.5390 & HIP29442~b & 18 \\
$4.70^{+1.50}_{-1.50}$ & 1.30$^{+0.06}_{-0.06}$ & 3.022 & TOI-2076~e & 19 \\
$4.72^{+0.42}_{-0.42}$ & 1.66$^{+0.04}_{-0.04}$ & 11.578 & HD136362b & 20 \\
$4.87^{+0.37}_{-0.37}$ & 1.53$^{+0.06}_{-0.06}$ & 1.2090 & GJ~9827~b & 21 \\
$4.97^{+0.24}_{-0.23}$ & 1.67$^{+0.17}_{-0.10}$ & 0.9596 & HD3167~b & 12 \\
$6.20^{+1.60}_{-1.60}$ & 2.69$^{+0.23}_{-0.23}$ & 22.891 & HD28109~b & 22 \\
$6.26^{+1.84}_{-1.83}$ & 1.84$^{+0.09}_{-0.09}$ & 1.0115 & WASP-132~c &23 \\
$6.48^{+0.99}_{-0.93}$ & 2.14$^{+0.26}_{-0.26}$ & 10.055 & K2-3~b & 21 \\
$6.70^{+2.80}_{-2.80}$ & 6.83$^{+0.13}_{-0.13}$ & 45.15396 & Kepler-51~b & 24 \\
$6.80^{+2.90}_{-2.90}$ & 1.85$^{+0.09}_{-0.09}$ & 4.78 & Kepler-48~b & 25 \\
$6.89^{+0.88}_{-0.88}$ & 2.60$^{+0.04}_{-0.04}$ & 15.572 & HIP~41378~b & 21 \\
$7.06^{+0.71}_{-0.68}$ & 1.79$^{+0.18}_{-0.06}$ & 0.790 & WASP-47~e &  21\\
$8.10^{+1.10}_{-1.10}$ & 3.01$^{+0.06}_{-0.06}$ & 4.3074 & TOI-1246b & 26 \\
$8.41^{+4.20}_{-4.20}$ & 1.51$^{+0.05}_{-0.05}$ & 1.5930 & Kepler-9~d & 25 \\
$8.56^{+1.54}_{-1.54}$ & 2.41$^{+0.05}_{-0.05}$ & 3.1275 & TOI-1260b & 27 \\
$8.75^{+1.09}_{-1.09}$ & 1.95$^{+0.09}_{-0.09}$ & 3.5951 & EPIC~249893012~b & 28 \\
$9.40^{+1.40}_{-1.40}$ & 2.31$^{+0.04}_{-0.04}$ & 3.390 & TOI-5624~b & 29\\
$9.50^{+0.90}_{-0.90}$ & 2.73$^{+0.08}_{-0.08}$ & 4.6538 & TOI-125~b & 30 \\
$9.60^{+0.90}_{-0.90}$ & 3.40$^{+0.08}_{-0.09}$ & 8.8803 & HD191939~b & 31 \\
$9.70^{+1.30}_{-1.30}$ & 1.77$^{+0.05}_{-0.03}$ & 3.696 & Kepler-20~b & 12 \\
$9.68^{+1.21}_{-1.37}$ & 2.59$^{+0.06}_{-0.06}$ & 3.471 & K2-285~b & 32 \\
$10.8^{+3.10}_{-3.10}$ & 2.70$^{+0.06}_{-0.06}$ & 6.2385 & Kepler-25~b & 25 \\
$11.0^{+1.29}_{-1.27}$ & 3.02$^{+0.08}_{-0.08}$ & 1.3484 & TOI-4010~b& 33\\
$12.8^{+3.90}_{-3.90}$ & 1.81$^{+0.21}_{-0.21}$ & 3.505 & Kepler-18~b & 25\\
$19.4^{+8.40}_{-8.40}$ & 3.70$^{+0.12}_{-0.12}$ & 13.748 & Kepler-92~b & 25 \\
$25.6^{+2.60}_{-2.60}$ & 2.40$^{+0.05}_{-0.05}$ & 3.005 & Kepler-411~b & 34 \\
$36.4^{+2.40}_{-2.40}$ & 4.77$^{+0.21}_{-0.21}$ & 10.501 & Kepler-56~b & 25 \\
\hline
\end{tabular}

{{\bf References:}  [1]  \cite{PBA23}; [2]  \cite{BDA22};  [3] \cite{Lel23}; 
 [4]  \cite{Lopez19};[5] \cite{LPD26}; [6] \cite{Luq23}; [7] \cite{Mills16};
 [8] \cite{SDH26}; [9] \cite{Agol21}; [10] \cite{KVG22}; [11] \cite{PZB24};
 [12] \cite{BDM23}; [13] \cite{WSCC26};  [14] \cite{GAS25}; [15]  \cite{HDT25}; 
 [16] \cite{BWH25}; [17] \cite{CBA24}; [18] \cite{EOK24}; [19]  \cite{WDL26};
 [20] \cite{DEA21}; [21] \cite{HSB25}; [22] \cite{BRG25}; [23] \cite{GBA25};
 [24] \cite{Masuda24}; [25] \cite{WIH24}; [26] \cite{TWD23};  [27] \cite{LCH23};
 [28] \cite{HPA20}; [29] \cite{BGLO26};  [30] \cite{NGA20}; [31] \cite{PLB24}; [32] \cite{PNL19};
 [33] \cite{KVH23}; [34] \cite{SIG19}}
 \end{table}

\begin{figure}
\centering
\includegraphics[width=1.0\linewidth]{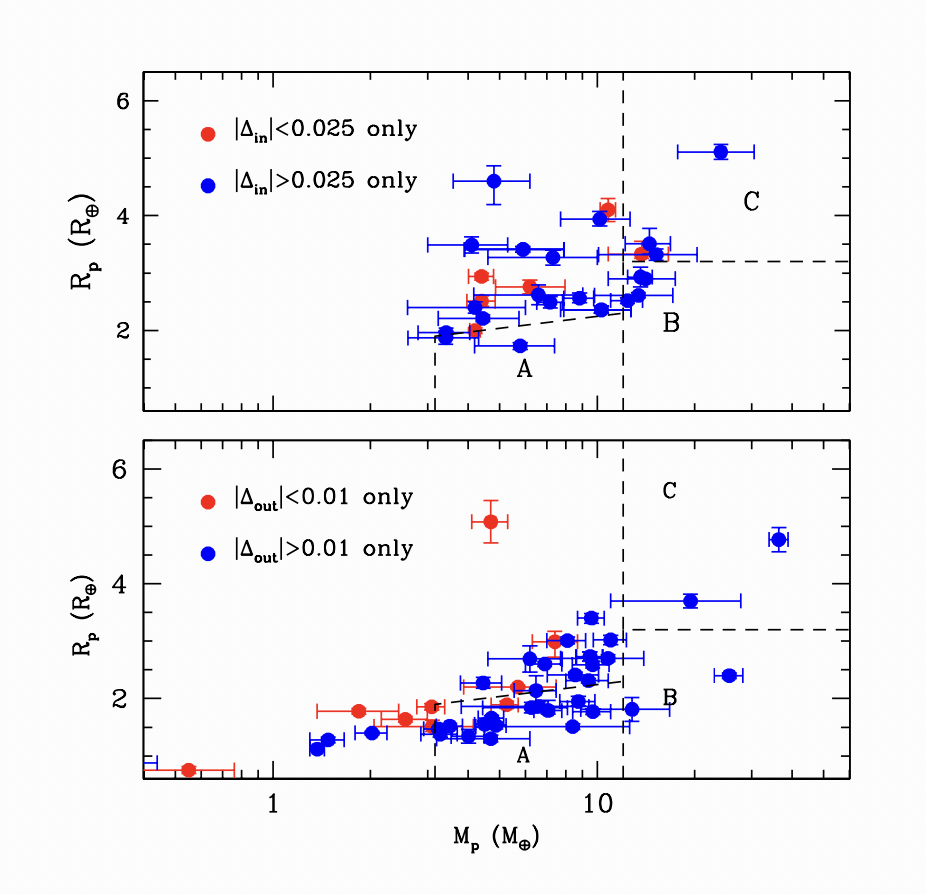}
\caption{The upper panel shows the properties of the outermost planets in our observed resonant chains. The
red points indicate planets that have likely to have not experienced a collision, while the blue points have
a probability of having experienced a collision. The lower panel shows the properties of the innermost planets
in the chains, with a similar colour scheme. In both panels, the regions labelled A, B and C are the same as 
before. \label{MRin}}
\end{figure}

As such, Figure~\ref{MRin} shows the mass--radius distributions for the outer planets (top panel)
and inner planets (lower panel) of our observed planetary chains. In each case, we divided the
samples into those that have a lower probability of collision (red points) and those which have a 
higher probability of collision (blue points). We make this distinction on the basis of the trends
shown in Figure~\ref{Dinout}. We designate a  planet as having a low probability of collision if
it has a  value of $\left| \Delta \right| $ such that  $>75\%$ of the corresponding planets in Figure~\ref{Dinout}
have not experienced a collision. 

For both inner and outer planets, we see that the planets with a low probability of  collision (red points) track the
same distribution as in Figure~\ref{MR1}.  For the outer planets, this largely holds true for the blue points as well, with
potentially an enhanced fraction in region~B. In the case of the inner planets, however, there is a substantial population
of planets in region~A. In accord with our discussion in \S~\ref{Unknown}, this is to be expected if region~A is populated
with planets whose properties are influenced by proximity to the central star and whose $\left| \Delta \right|$ is increased
by a process such as tidal circularisation or photoevaporation.

\begin{table}[ht]
\caption{Outermost planets of chains ($M<100 M_{\oplus}$) \label{OuterMR}}
\begin{tabular}{|lllll|}
\hline
$M_p$  & $R_p$  & Period  & Name & Ref. \\
 ($M_{\oplus}$) & ($R_{\oplus}$)  & (days) &   &\\
\hline
\multicolumn{5}{c}{Outer Planets with $\left| \Delta_{in} <0.025 \right|$} \\
\hline
$0.33^{+0.02}_{-0.02}$ & 0.76$^{+0.01}_{-0.01}$ & 18.772 & Trappist-1h & 1\\
$4.20^{+0.16}_{-0.16}$ & 2.00$^{+0.05}_{-0.05}$ & 11.3819 & TOI-270~d & 2\\
$4.40^{+0.44}_{-0.44}$ & 2.51$^{+0.09}_{-0.09}$ & 11.90 & Kepler-60~d & 3\\
$4.40^{+0.39}_{-0.39}$ & 2.94$^{+0.06}_{-0.06}$ & 20.717 & TOI-178~g & 4 \\
$6.20^{+1.76}_{-1.34}$ & 2.76$^{+0.12}_{-0.12}$ & 16.74 & Kepler-305~d & 3 \\
$10.8^{+0.60}_{-0.60}$ & 4.1$^{+0.2}_{-0.2}$ & 11.899 & K2-19~c & 5\\
$12.0^{+3.00}_{-3.00}$ & 9.20$^{+0.10}_{-0.10}$ & 542.08 & HIP 41378~f & 5 \\
$13.7^{+2.90}_{-2.90}$ & 3.33$^{+0.22}_{-0.22}$ & 31.29 & K2-32~d & 5\\
$15.3^{+4.20}_{-4.20}$ & 10.17$^{+0.75}_{-0.75}$ & 48.67 & V1298~Tau~e & 6 \\
$29.9^{+1.10}_{-1.30}$ & 8.08$^{+0.53}_{-0.41}$ & 38.985 & Kepler-9~c & 7 \\
\hline
\multicolumn{5}{c}{Outer Planets with $\left| \Delta_{in} >0.025 \right|$} \\
\hline
$3.41^{+0.90}_{-0.90}$ & 1.87$^{+0.12}_{-0.11}$ & 29.797 & HD23472~c & 8 \\
$3.42^{+0.62}_{-0.62}$ & 1.96$^{+0.08}_{-0.08}$ & 6.2018 & GJ~9827~d & 5 \\
$4.10^{+1.20}_{-1.10}$ & 3.49$^{+0.14}_{-0.14}$ & 81.066 & Kepler-79~e & 9\\
$4.20^{+1.80}_{-1.60}$ & 2.40$^{+0.10}_{-0.10}$ & 138.07 & HD224018~d & 10\\
$4.44^{+1.30}_{-1.21}$ & 2.21$^{+0.06}_{-0.06}$ & 15.27 & Kepler-23~d & 3 \\
$4.80^{+1.40}_{-1.20}$ & 4.60$^{+0.27}_{-0.41}$ & 19.726 & Kepler-223~d & 11 \\
$5.79^{+1.60}_{-1.60}$ & 1.73$^{+0.06}_{-0.06}$ & 29.318 & LHS~1903~e & 12 \\
$5.90^{+2.00}_{-2.00}$ & 3.41$^{+0.05}_{-0.05}$ & 13.6308 & TOI-469.01 & 13 \\
$6.59^{+2.43}_{-2.43}$ & 2.62$^{+0.17}_{-0.17}$ & 43.844 & Kepler-106~e & 14 \\
$7.16^{+0.65}_{-0.65}$ & 2.49$^{+0.10}_{-0.10}$ & 4.9899 & LP791-18~d &  15\\
$7.30^{+2.70}_{-2.70}$ & 3.27$^{+0.13}_{-0.13}$ & 35.13 & TOI-2076~d & 16 \\
$8.82^{+0.93}_{-0.92}$ & 2.56$^{+0.09}_{-0.08}$ & 107.25 & HD136352~d & 17 \\
$10.18^{+2.46}_{-2.46}$ & 3.94$^{+0.13}_{-0.13}$ & 35.747 & EPIC249893012~d & 18 \\
$10.3^{+2.40}_{-2.60}$ & 2.363$^{+0.06}_{-0.06}$ & 24.3681 & K2-233~d  & 3 \\
$12.4^{+1.40}_{-1.40}$ & 2.52$^{+0.05}_{-0.05}$ & 77.144 & TOI-561~e & 19 \\
$13.4^{+3.70}_{-3.40}$ & 2.61$^{+0.05}_{-0.04}$ & 77.611 & Kepler-20~d & 20\\
$13.6^{+1.20}_{-1.20}$ & 2.93$^{+0.17}_{-0.17}$ & 19.98 & TOI-125~d & 21\\
$14.1^{+3.30}_{-3.30}$ & 2.90$^{+0.04}_{-0.04}$ & 14.749 & Kepler-107~e & 20 \\
$14.5^{+2.30}_{-2.30}$ & 3.51$^{+0.27}_{-0.18}$ & 37.925 & TOI-1246e & 22 \\
$15.2^{+5.10}_{-5.10}$ & 3.32$^{+0.10}_{-0.10}$ & 58.020 & Kepler-411~d & 23 \\
$24.6^{+6.40}_{-6.40}$ & 5.11$^{+0.13}_{-0.13}$  & 14.859 & Kepler-18d & 14\\
\hline
\end{tabular}

{{\bf References:} [1] \cite{Agol21}; [2]   \cite{KVG22}; [3]  \cite{Lel23}; [4]  \cite{SDH26};
[5] \cite{HSB25}; [6] \cite{LPD26}; [7] \cite{BMP19}; [8]\cite{BDA22}; [9] \cite{JHL14}; [10]  \cite{DNA25};
[11] \cite{Mills16};  [12]  \cite{WSCC26};[13] \cite{EOK24}; [14] \cite{WIH24}; [15] \cite{GMV25}; 
[16] \cite{WDL26}; [17] \cite{DEA21}; [18] \cite{HPA20}; [19]  \cite{GBA25}; [20] \cite{BDM23}; [21] \cite{NGA20}; [22] \cite{PLB24};
[23] \cite{SIG19}}
\end{table}

\subsection{Mass loss via Evaporation}

Implicit in our original hypothesis was the notion that planets  lose their volatile envelopes due to collisions  during the giant impact phase that results
from dynamical instability. However, this is not the only way to lose a volatile envelope. It is also
possible that planets can lose envelope mass through evaporation, powered either by photoevaporation by the central star \citep{OW13,LF13}
or by residual thermal heat of formation \citep{GSS18,GS19}. Such mass loss would have a slightly different signature in the mass--radius plot,
since it would reduce the radius without appreciable change in the mass (estimated envelope mass fractions are mostly a few percent). 
Giant impacts, on the other hand, usually increase the mass
of the remnant, while also reducing the low density component.

\begin{figure}
\centering
\includegraphics[width=1.0\linewidth]{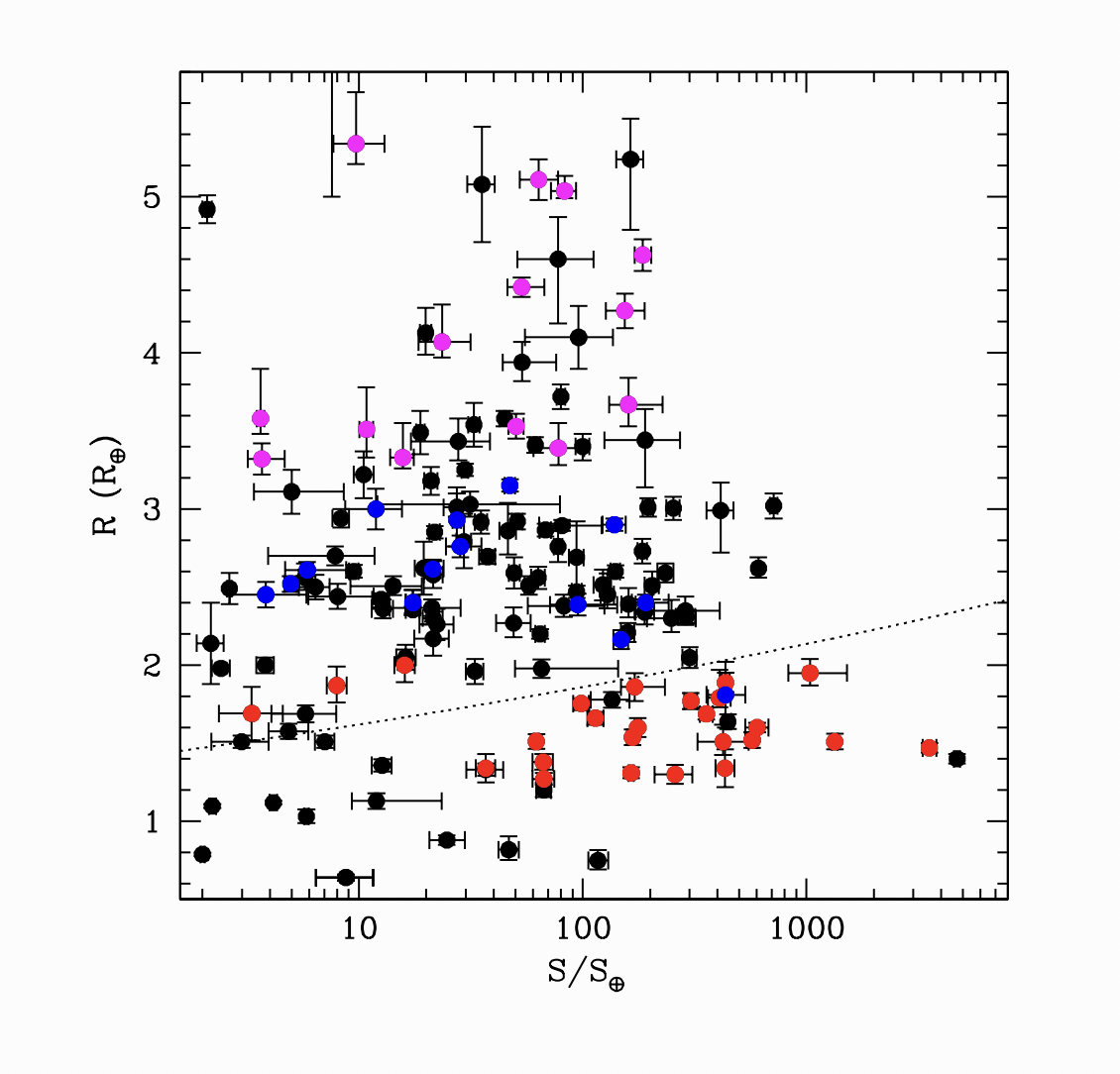}
\caption{Each point shows a radius of a planet in our sample, plotted against the flux incident on the planet from
the host star (measured in units of Earth's Solar irradiance). The sample shown here is the cumulative sum of Tables~\ref{PristineMR}, \ref{CollProducts},
\ref{MuddyMR}, \ref{InnerMR} and \ref{OuterMR}.
The red points indicate those points that fall within box A and the blue points those that fall within box B.  The magenta
points fall in region~C. The
dashed line shows the fit to the radius valley as a function of irradiation from \cite{PRI22}. \label{RSplot}}
\end{figure}

Figure~\ref{RSplot} shows the radii of  our planetary sample as a function of the irradiation they receive from their host star, measured in units of the Solar irradiance. The red points indicate the 
planets in Region~A (from all the  populations in Tables 1--5).  There is clearly a strong bias towards higher irradiation in this population, which supports the idea that these rocky planets
began with low density envelopes, but lost them via evaporative processes. Not surprisingly, a significant fraction of this group are drawn from the innermost members
of planetary chains. The dotted line in Figure~\ref{RSplot} shows the location of the `radius valley' \citep{PRI22} as a function of irradiation -- indicating that the distribution
seen here is consistent with the broader population as well.

The blue points show the planets that are found in Region~B. There is no similar bias towards higher irradiation in this group, so this group is more plausibly associated
with an origin stemming from the dynamical evolution of the planetary systems. Similarly, the magenta points show the distribution of planets from region~C, which
also show no strong correlation with the level of irradiation.

We therefore conclude that most of the planets that population region~A are more plausibly explained by evaporation of low density atmospheres, instead of removal during
late stage giant impacts. On the basis of these comparisons it appears as though most rocky planets with mass $> 3 M_{\oplus}$ are the remnants of planets that
began with volatile envelopes but lost them. Rocky planets of mass $< 3 M_{\oplus}$ do not share this bias to high irradiation, and are plausibly primordial.


\section{Discussion}
\label{Infer}

Dynamical evolution of a planetary system can have an effect on both the system architecture and the instrinsic planetary
properties themselves. By examining the correlations between these two consequences, we can hope to test different models
for planetary system formation and evolution.

\subsection{Bulk  Masses}

\begin{figure}
\centering
\includegraphics[width=1.0\linewidth]{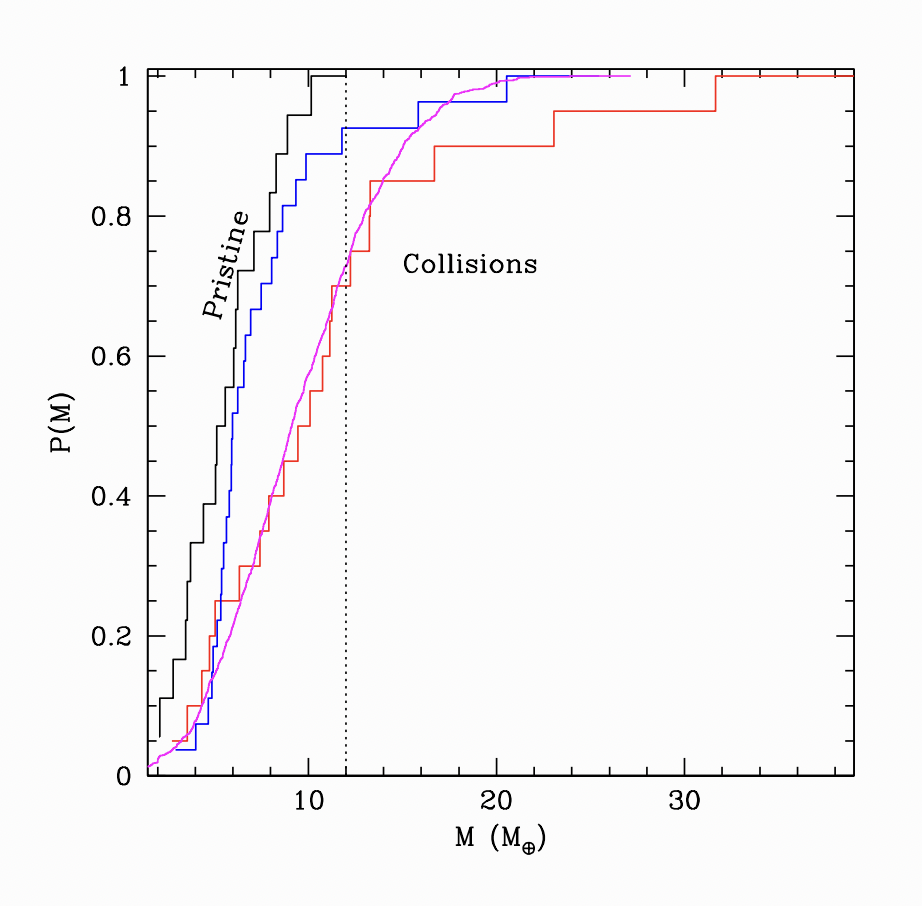}
\caption{The black histogram represents the mass distribution of the Pristine planets, which contains no planets with masses $> 12 M_{\oplus}$,
a value indicated by the vertical dotted line. The red histogram is the mass distribution of the possible collisional products, including those in
region~B, but not in region~C. The blue histogram represents the mass distribution of the unassigned planets, again including region~B but not region~C.
The magenta histogram is the mass distribution obtained by combining the masses of two planets from the Pristine population, chosen at
random.
 \label{NMass}}
\end{figure}

Examination of the distribution of planet properties in Figures~\ref{MR1} and \ref{MR2} does suggest a shift towards larger masses
in the sub-Neptune groups, when going from pristine to collisional populations. As a test of this impression, Figure~\ref{NMass} shows
the cumulative distribution of mass for the sub-Neptune populations (defined as those with radii $> 2 R_{\oplus}$, but not including
those in region~C) in Figures~\ref{MR1}, \ref{MR2} and \ref{MR3}.

The distributions in Figure~\ref{NMass} do indeed support the hypothesis that the collisional remnants are more massive than the pristine planets
(as noted by \cite{,LDB24} and \cite{LCCD25}). The mass distribution of the unassigned planets suggests that most of this population are also pristine.

Is this mass distribution plausibly the outcome of collisional accumulation? The magenta histogram in Figure~\ref{NMass} is derived by
sampling the full mass distribution of the pristine planets (the black histogram) and combining two planets, randomly chosen, to make a collision product. Each
mass is sampled within its 1$\sigma$ error bar and recorded. The fact that the magenta histogram is a plausible match to the red histogram
indicates that the collisional population is indeed consistent with being drawn from the dynamical dissolution and collisional accumulation
of the pristine planets. The most massive planets are underrepresented in the magenta sample, but may represent the small fraction of planets
that undergo multiple mergers.

It should be noted that our samples are drawn from a heterogeneous set of surveys, so estimates of occurrence rates based on this data
are unreliable, but the data are consistent with the model that the population shown in Figure~\ref{MR2} represents collision products
drawn from the sample in Figure~\ref{MR1}.

\subsection{Volatile Mass Fractions}

One potential complication, however, is that many of these putative collision remnants still seem to possess significant Hydrogen atmospheres. 
To obtain the Hydrogen mass fraction $f_H$ for each planet, we interpolate the mass, radius and insolation for each planet, in the tables
of \cite{LF14}. Given the traditional age uncertainties for main sequence stars, we use both the 1~Gyr and 10~Gyr old tables, and use the
difference in results as an estimate in the likely error of each determination. Upper limits are given if the measured parameters fall below
the model radius grid for either of the two ages. If it falls outside the grid for both ages, the upper limit is given as $f_H=0.01 \%$, as this
is the lowest value in the model grid.

\begin{figure}
\centering
\includegraphics[width=1.0\linewidth]{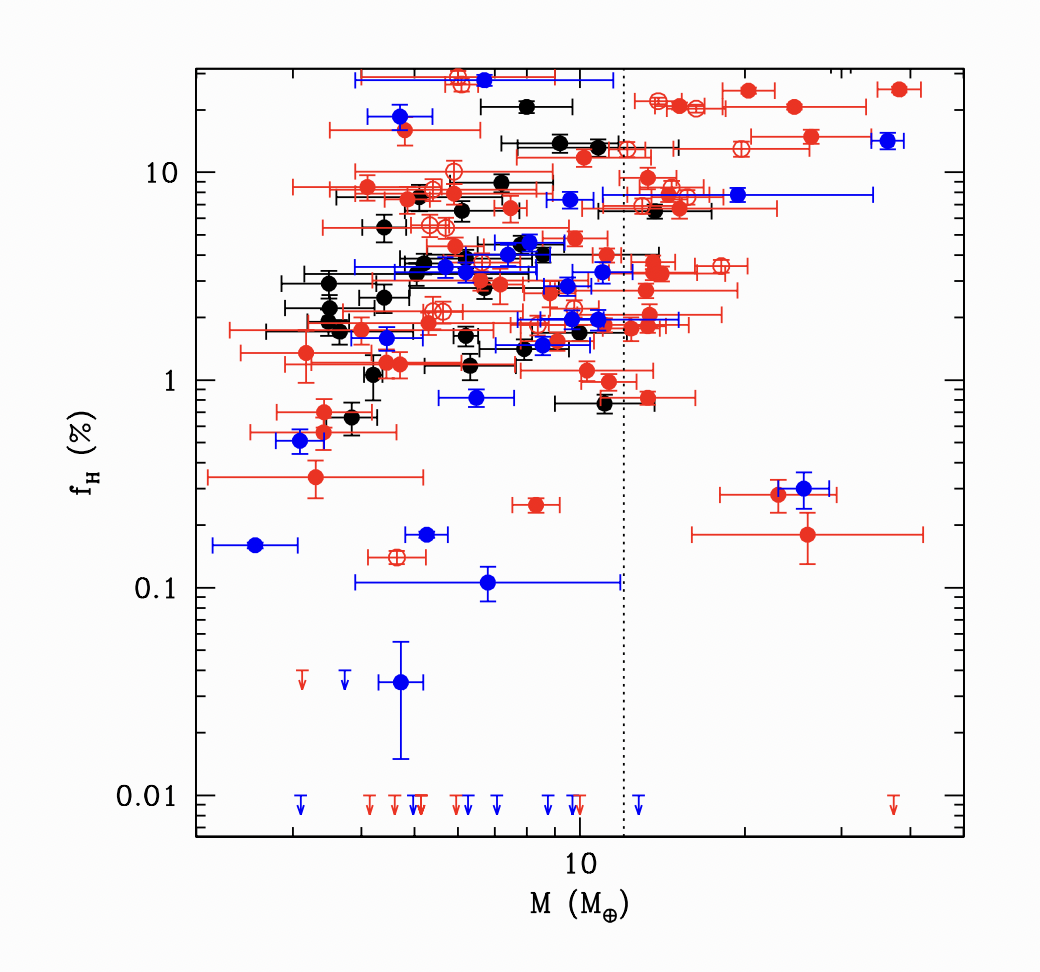}
\caption{The  black points represent the Hydrogen atmosphere mass fractions, as a function of mass, for the pristine population from Tables~\ref{PristineMR}
and \ref{OuterMR}. The red points represent the corresponding mass fractions for the collision products from Tables~\ref{CollProducts} and \ref{OuterMR}.
Open red points are from the unassigned planets in Table~\ref{MuddyMR}.
The vertical dotted line represents the mass boundary ($12 M_{\oplus}$) of region~A. The blue points represent the inner planets, from Table~\ref{InnerMR},
whose Hydrogen mass fractions are more likely to have been affected by evaporation.  The extension of the distribution towards the upper
right represents the contribution from planets in Region~C, which likely have a different formation history than the lower mass planets.
Planets shown as upper limits are consistent with bare, rocky cores. \label{fHM}}
\end{figure}

We see that most pristine planets with masses $>3 M_{\oplus}$ have H mass fractions in the range 1--10\%. 
 The shift in planetary masses between the two populations shown in Figures~\ref{MR1} and \ref{MR2} suggests impactor/target mass ratios mostly close to
unity. This is especially true for the planets in Region~B, as they represent a sample not found in the pristine sample and thus must have
acquired significant additional mass.  However, as shown in Figure~\ref{fHM}, the Hydrogen mass fractions of the nominal collision products  appear to be very similar to
those of the pristine planets.
Only for the most massive planets ($> 20 M_{\oplus}$) does there seem to be a population with a lower H mass fraction, which is only
a small fraction of the overall population. 

Planetary
scale giant impacts are expected to eventually remove nearly all the volatile atmosphere component unless the impactor mass is $<10\%$ the mass of
the planet \citep{BS19}, and the shift in planetary masses between the two populations suggests impactor/target mass ratios mostly close to
unity. This is especially true for the planets in Region~B, as they represent a sample not found in the pristine sample and thus must have
acquired significant additional mass.  

\subsection{Sub-Saturns}

We have noted that the planets with mass $> 12 M_{\oplus}$ appear to be bimodal, with drastically different Hydrogen mass
fractions. Those in region~B are plausibly related to the lower mass planets in the pristine sample. However, those in Region~C
have radii that are too large to be explained by Hydrogen mass fractions in the range explored in Figure~\ref{fHM}. 

Thus, the planets in Region~C are likely to represent a different mode of planet formation (the `Sub-Saturns')  from the one we describe
here. This is supported by the observation that the occurrence rates of planets drops substantially above a threshold
radius $\sim 3.5 R_{\oplus}$ \citep{FPH17,GPE25}, suggesting a transition to a different mode of planet formation. Indeed,
several analyses conclude that more than one formation pathway may be applicable in this mass range \citep{FDH25,GPE25,TNAC25},
given the range of eccentricities, envelope mass fractions and multiplicities observed. 

\subsection{Primordial Mass Range}

\begin{figure}
\centering
\includegraphics[width=1.0\linewidth]{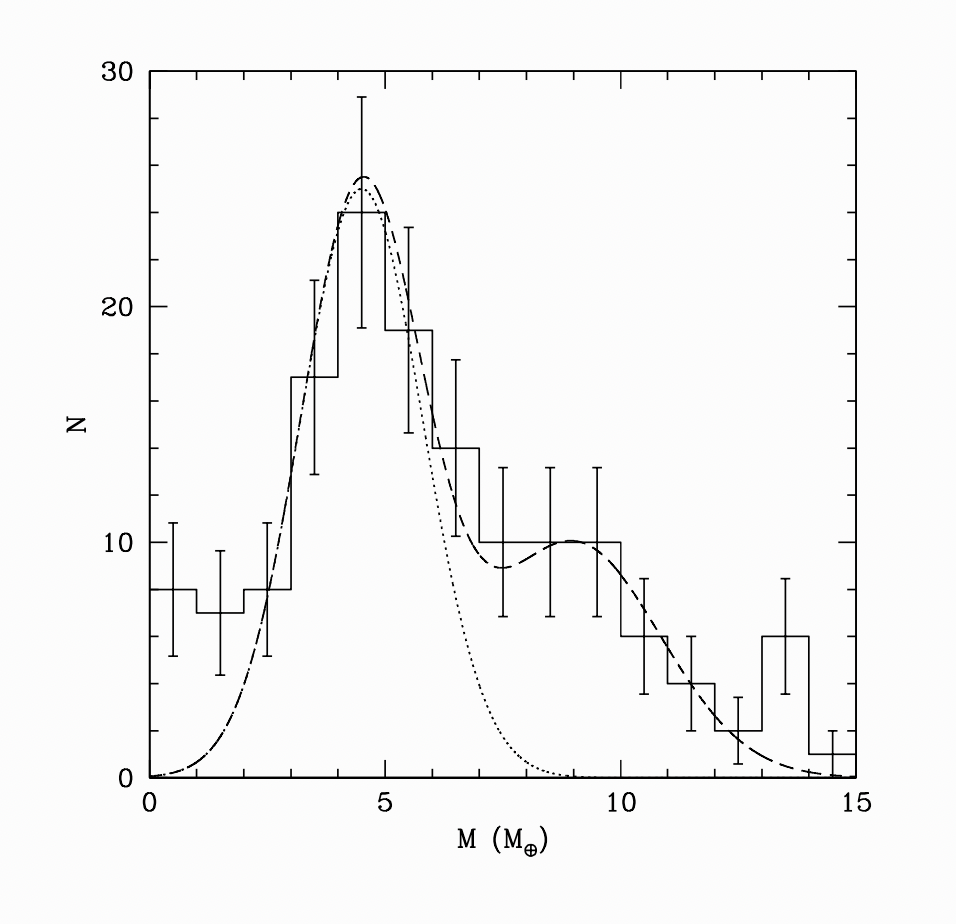}
\caption{The solid histogram represents the mass function of planets  from Tables~\ref{PristineMR}, \ref{CollProducts}, \ref{MuddyMR},
\ref{InnerMR} and \ref{OuterMR}, excepting those massive planets that lie in region~C -- which we assert must have formed from
a different mechanism (see discussion in the text). The dotted histogram represents a gaussian distribution centered at 4.5 $M_{\oplus}$,
with a dispersion of $1.3 M_{\oplus}$. The dashed histogram arises if we add a contribution of
equivalent total mass by randomly combining pairs from the dotted histogram together into single objects. The peak at
13 $M_{\oplus}$ is also equivalent to adding three planets from the pristine distribution together.
 The sample here contains
more planets than the ones in Figure~\ref{NMass}, since we have included the planets in region~A.
 \label{Mbin}}
\end{figure}

Visual inspection of Figures~\ref{MR1}, \ref{MR3} or \ref{NMass} suggests that many of the planets
in our sample have masses in the range 5--7 $M_{\odot}$. To investigate this further, Figure~\ref{Mbin} shows
the mass function of all planets in our sample, except for those that fall into region~C. We include here the
planets in region~A because, even if they have lost their primordial Hydrogen atmospheres, the original mass fraction of the lost envelope 
is much less than the $1 M_{\oplus}$ size of the mass bins in this plot.

This mass function shows a pronounced spike at masses $\sim 4$--6$M_{\oplus}$, suggesting a characteristic mass
associated with this population. This is consistent with contemporary models for the origin of the migrating planets,
wherein masses $\sim 5 M_{\oplus}$ are required to break out of the original birth locations at planet traps on
scales of $> 1 AU$ and migrate inwards \citep[e.g.][]{Has16,IOR17}.  The distribution in Figure~\ref{Mbin} also
shows a minor component at lower masses, which are consistent with forming without substantial Hydrogen
atmospheres. There is also a tail towards larger masses which are consistent with the formation through
collisions. To demonstrate this we have fit the peak with a Gaussian distribution of masses, centered at 4.5 $M_{\oplus}$,
with a dispersion of 1.3 $M_{\oplus}$. To simulate a collisional population, we randomly sample pairs
from this distribution and add them. If we add the resulting masses to the original distribution (in a 1:1 ratio) we find the
dashed histogram. Although this sample of planets is derived from a heterogenous set of observations, so that the
relative proportions may be uncertain, the correspondence
between the model and the data does indicate the plausibility of the model that this sample of planets is born from a
narrow mass distribution and broadened by collisional mergers.

\section{Conclusions}

The conclusions we draw from Figures~\ref{NMass} and \ref{fHM} seem almost contradictory. The distribution
of bulk masses between the pristine planets and the collision products supports the notion that the latter population
contains a substantial contribution from mergers amongst members of the former population. However, the
Hydrogen mass fractions do not seem to change substantially between the two populations, despite the fact that
the energy input from the collisions is expected to largely unbind the envelope.

The best way to reconcile these two observations is to adopt a model in which the collisions occur early enough
that some fraction of  the protoplanetary disk is still present, so that the collisional products have a chance to
recapture gas from the dissipating nebula. A rocky core, embedded in a gaseous disk, establishes an equilibrium
envelope mass regulated  by the interaction of the planetary potential well, gaseous pressure, and cooling \citep{RRR06,LCO14,LC15,GSS16}.
However, the final size of  remnant atmosphere is determined
largely during the disk dispersal phase \citep{ROS24}, as the exposure to the stellar irradiation `boils off' much
of the Hydrogen originally  bound to the planet during the disk phase. The fact that dynamically pristine planets and
collision products seem to retain similar atmospheric Hydrogen budgets indicates that they both established their
atmospheres before these latter stages and therefore experienced similar boil-off phases. Therefore, the collisions must have occurred before the final dispersal
of the protoplanetary disk.

Further support for this timeline comes from the observation that 
  planets
with $M<2 M_{\oplus}$ seem to form with little H envelope, while those with larger mass cores acquire a H envelope
of a few percent, with the transition mass in the range 2--3 $M_{\oplus}$.  Our results suggest that rocky planets
with observed masses $> 3 M_{\oplus}$ are largely the cores of planets whose envelopes evaporated. In principle,
such planets could also be formed in collisions, but our results above suggest that most collisions occur early enough
that the collision products re-accrete their envelopes.  A lower limit of 2--3 $M_{\oplus}$ for planets with
significant Hydrogen atmospheres is also consistent with the limit expected for planets whose envelopes
are set during the late stages of protoplanetary disk evolution \citep{LKT22}. Planets with masses $< 2 M_{\oplus}$
are expected to retain very small Hydrogen envelopes under the conditions late in the disk lifetime, where the
atmospheres are effectively isothermal.

This conclusion has direct implications for models of the dynamical evolution of compact planetary systems.
The gravitational interactions with the protoplanetary disk drive the inward migration of the planetary systems,
leading to the establishment of resonant chains.
In most models, this takes place rapidly ($<10^5$ years), so that the original establishment of the chains
is not an impediment to the presence of gas near the planets.  The central question is how the chains
are broken.

In the popular `breaking-the-chains model' \citep{IOR17,IBR21} the resonant chains survive to the end
of the protoplanetary disk phase, and only become dynamically unstable when the eccentricity damping
from the disk goes away. In this model, 
the bulk of the dynamical evolution occurs after
the disk has dissipated, so that the collisions are expected to occur in the absence of gas. This model
seems to be contradicted by the similarity in the Hydrogen mass fractions observed here.

Magnetospheric rebound models offer a more promising avenue. In this class of models, the outward
motion of the magnetospheric cavity, as the disk ram pressure wanes, drives additional dynamical evolution.
Simple, sharp-edged, models of magnetospheric
cavity evolution \citep{LOL17,LO17}  have difficulty generating enough instability, but models with more
physically motivated inner edges \citep{YHH23}  can drive enough dynamical evolution to break most of the resonant chains \citep{HYH24,HYNH25}
during the final disk dispersal phase.
These models do not generate as many collisions as in the `breaking-the chains' models, but dynamical instability does still
occur a non-negligible fraction of the time. In this case, many of the collisions occur early enough that the collision products can still acquire gas
from the nebula after the collision.

A third scenario, termed `rattle-and-hum', invokes a population of remnant planetesimals to scatter off of surviving 
planets to drive them away from resonance \citep{CF15,WML24}. In this model, the scattering also generates finite
free eccentricity and can lead to dynamical instability of the system on longer timescales. Recent estimates \citep{HW26}
suggest that the scatterers should be of order a Mercury mass to generate the observed levels of eccentricity, but that
such masses drive evolution on too fast of a timescale. To match the claim (see below) that the dynamical instability
timescale is $\sim 100$~Myr, the scattering bodies should be of order Pluto's mass instead. Our results here can potentially
resolve this conundrum, as a more rapid evolution would imply masses consistent with those needed to match the observed
free eccentricities. Of course, the source of the remnant planetesimal population in this model is still unknown. The
influence of the accreted high-mean-molecular-weight material on the inferred Hydrogen layer thicknesses is also
something that remains to be calculated.

There remain a handful of uncertain issues worthy of further study in this context.
\begin{itemize}
\item {\bf Observed timescale of genuine dynamical instability:}
The conclusion that collisions must occur early enough for gas to still be present seems to conflict with recent claims that the resonant fraction of planetary
systems evolves considerably on timescales $\sim 100$~Myr \citep{DGB24}. However, such claims should be interpreted
with some caution. The high resonant fraction of the young (10 Myr) sample in \cite{DGB24} is of limited statistical significance.
Of the 10 observed pairs, two contain a planet 
that was discovered via its TTV influence on the neighbour. Since TTV variations are
stronger when a pair is close to resonance, using such systems to infer a high resonant fraction is circular. Removal of the biased pairs reduces the statistical
basis of the claim to the point that no claim can be made. Furthermore, the implication that these systems represent a sample of planets
emerging from the disk phase primed to undergo dynamical instability is false. The nominal resonant chains that emerge from the disk phase
in the breaking-the-chains scenario are closely packed, with many pairs in 4:3, 5:4 and 6:5 resonances (as illustrated in
Figure~15 of \cite{WMP23}). The pairs observed by \cite{DGB24} are
locked in 3:2 and 2:1 resonances, and some can actually be demonstrated to not actually be in resonance or likely to
go unstable without an external excitation of eccentricity \citep{HDZ25}. 

In fact, the observed young sample of \cite{DGB24} may represent the outcomes of an initial phase of collisional and dynamical evolution
that took place within the protoplanetary disk phase, awaiting additional dynamical evolution due to external secular perturbations from planetary
or stellar companions \citep{HA16,H17,HPD17,BA17,MDJ17,PL19,DNH19,PN20,PL21}. In such a scenario, the physical properties of the planets would comport with the results presented here.
Unfortunately, there is only one system in the \cite{DGB24} young sample that meets the criteria for inclusion in our sample -- V1298~Tau.
Planet V1298~Tau~d is consistent with being a pristine planet, while V1298~Tau~b is unassigned.

There is limited evidence that the planets in Table~\ref{PristineMR} are younger than the average, in the sense that, of
the planets with $M<10 M_{\oplus}$ and $R>4 R_{\oplus}$, four are found in Table~\ref{PristineMR}, none in
Table~\ref{CollProducts} and two in Table~\ref{MuddyMR}. However, only V1298~Tau~d and the planets
Kepler-51 c/d orbit stars with ages $< 1$Gyr. This is a small fraction of the total population, so our sample also contains
little direct support for the idea that the properties of the planets evolve over Gyr timescales.

\item {\bf Trends with Central Star Mass:} The sample of planets surveyed here orbits a range of stellar host masses ranging 
from $\sim 0.09 M_{\odot}$ (Trappist-1) to $\sim 1.25 M_{\odot}$ (Kepler-65~c). With the limited statistical significance of a small
sample, we have assumed that planetary composition does not depend on the host star properties (mass or chemical composition).
This is obviously a simplification, and may contribute to some of the dispersion in planetary properties (such as the size of
the surviving volatile envelope).

\begin{figure}
\centering
\includegraphics[width=1.0\linewidth]{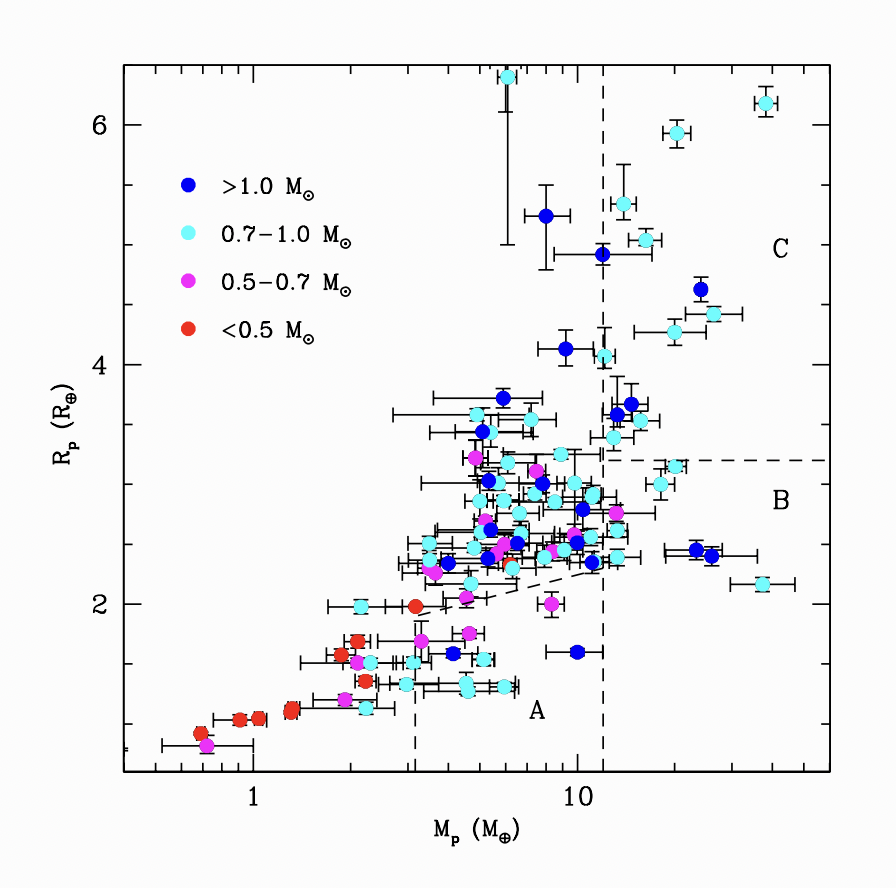}
\caption{The  red points show planets that orbit M dwarf ($M_*<0.5 M_{\odot}$) hosts. The magenta points
orbit K dwarf ($0.5$--$0.7 M_{\odot}$) hosts. The cyan points orbit G dwarfs ($0.7$--$1.0 M_{\odot}$) and
the blue points orbit F or A stars ($M>1.0 M_{\odot}$).  \label{LLM}}
\end{figure}

Figure~\ref{LLM} shows the properties of planets in Table~\ref{PristineMR}, Table~\ref{CollProducts} and Table~\ref{MuddyMR}, as a function of host star mass.
We see that the low mass planets are mostly orbiting M dwarfs (dominated by the sample from the Trappist-1 system)
while the planets with the largest radii orbit F stars (a likely indicator of youth). The bulk of the sample orbit G stars and
both the pristine and the collision products are predominantly drawn from this group, so there is little indication that the
difference is driven by stellar host properties.
The predominance of low mass hosts for planets with $M<3 M_{\oplus}$ is likely to be a selection effect, as smaller and
lighter stellar hosts allow for the detection of lower mass planets. 

There is also little difference in host star metallicity between the pristine and collisional groups -- a KS test indicates that
the host star metalliicity distributions for the two samples have a 69\% chance of being drawn from the same underlying distribution.

\item {\bf Merging, mass loss and reaccretion:} We have assumed that giant planetary impacts result in the loss of the volatile
envelopes but the perfect merging of the rocky cores. Simulations of direct, impact-generated mass loss \citep[e.g.][]{GA03,RLD25} show
a variety of outcomes that depend on the impact details. However, the longer term consequences of impact heating drive evaporation
of low density atmospheric components \citep{BS19}.
The temperatures of the evaporating 
Hydrogen gas heated by a collision will result in ionization and coupling to a stellar wind. Indeed, the fate of the gas -- once it has
been lifted free of the potential well of the planet -- should be similar whether it was removed in a collision or by photoevaporation, as the
underlying physics is very similar \citep{GSS18,GS19}
and the population of planets in Region~A demonstrate that such gas is indeed lost. The assumption of perfect merging of the rocky bodies
 is less clear, and may be inaccurate for extended collisional cascades \citep{HM22} but is a reasonable approximation for individual
 collisions of planetary scale bodies \citep{KG10}. In  part this is because 
 debris released in a planetary collision is
largely re-accreted by the original body, since the planetary escape velocities are much less than the escape velocities from the stellar potential well.
\item {\bf Hydrogen sequestration in the core:}
We have drawn a simple division between rocky cores and Hydrogen envelopes, but recent work on the 
chemical reactions between magma oceans and Hydrogen envelopes \citep{CS18,KFS20,GSS25,RYS25} could muddy the distinction. 
Hydrogen sequestered in the core could potentially outgas at later times and replace an atmosphere. The
degree to which such sequestered material could survive a collision is also unclear, since much of the 
atmosphere lost in collision is actually due to the heating of the core itself \citep{BS19}. 
\end{itemize}

Taken in total, our results confirm previous claims that planets presently close to resonance today have, on average, lower masses
than those whose orbital architectures suggest a history of collisional evolution. However, we find little difference in the Hydrogen
envelope mass fraction between these two populations. This leads us to conclude that the collisional evolution in multiple planet systems
must occur early, so that the products of large scale planetary collisions can re-accrete gas from the primordial nebula. 
On the other hand, the mass distribution we infer for the pristine planets is in good accord with formation models for sub-Neptune
planets  that produce
$\sim 5 M_{\oplus}$ cores which grow in the outer disk and then migrate inwards when they reach a threshold mass. Rocky
planets of smaller ($< 3 M_{\oplus}$) also form, but rocky planets observed with larger masses seem primarily to be the
product of photoevaporation of primordial Hydrogen atmospheres.

Our conclusions bear several similarities to those of \cite{HHB26}, which was based on a larger sample but did not include
dynamical architecture information.  Although we see little evidence for super-Mercuries in our sample, we too find that the
rocky planets divide into primordial cores at low masses and photoevaporated cores at higher masses.
For planets with low density envelopes,
our classifications are qualitatively similar, although we find the role of collisions to be less than inferred by \cite{HHB26},
based on the fact that our pristine sample overlaps significantly with the parameter space  of their collisionally sculpted populations.

In conclusion, our results are in broad agreement with the contemporary picture of how compact planetary systems form, with
one exception -- most of the dynamical instability and resulting planetary collisions must occur early in the lifetime of the planetary
system -- when there was still gas present in the protoplanetary disk.

{\bf Data availability}: The data underlying this article will be shared on reasonable request to the corresponding author.

This research has made use of NASA's Astrophysics Data System Bibliographic Services. 
This research has made use of the NASA Exoplanet Archive, which is operated by the California Institute of Technology, under contract with the National Aeronautics and Space Administration under the Exoplanet Exploration Program. This research has made use of NASA's Astrophysics Data System Bibliographic Services. 
The simulations described here were performed on the UCLA Hoffman2 Shared computing cluster.

\bibliographystyle{mnras}
\bibliography{MRrefs}

\end{document}